 \documentclass[twocol]{ametsocV6.1}

\newcommand{\calcof}{CalCOFI}

\newcommand{\dounit}{\ensuremath{\mu {\rm mol \, kg^{-1}} }}
\newcommand{\denunits}{\ensuremath{{\rm kg \, m^{-3}}}}
\newcommand{\salunits}{\ensuremath{{\rm g \, kg^{-1}}}}
\newcommand{\nunits}{\ensuremath{{\rm cycles \, hr^{-1}}}}
\newcommand{\degc}{\ensuremath{{\rm deg \, C}}}
\newcommand{\chlunits}{\ensuremath{{\rm mg \, m^{-3}}}}

\newcommand{\abssal}{\ensuremath{S_A}}
\newcommand{\const}{\ensuremath{\theta}}
\newcommand{\potd}{\ensuremath{\sigma_0}}

\newcommand{\pjasd}{\ensuremath{p(\abssal,\potd)}}
\newcommand{\pjson}{\ensuremath{p(\so,\buoy)}}
\newcommand{\pjtdo}{\ensuremath{p(\const,\doxy)}}

\newcommand{\doxy}{\ensuremath{{\rm DO}}}
\newcommand{\buoy}{\ensuremath{N}}
\newcommand{\chla}{\ensuremath{{\rm Chl-a}}}

\newcommand{\oc}{\ensuremath{{\rm OC}}} 
\newcommand{\so}{\ensuremath{{\rm SO}}} 

\newcommand{\exso}{\ensuremath{{\rm SO_{\rm U}}}} 
\newcommand{\mnso}{\ensuremath{{\rm SO_{\rm L}}}} 
\newcommand{\valexso}{1.1}
\newcommand{\events}{events}  
\newcommand{\event}{event}  

\newcommand{\cugn}{CUGN}

\title{Extremes of Dissolved Oxygen in the California Current System}

\authors{J. Xavier Prochaska, \aff{a,b,c,d}\correspondingauthor{J. Xavier Prochaska, jxp@ucsc.edu} 
Daniel Rudnick,\aff{e} 
}

\affiliation{\aff{a}{Affiliate of the Department of Ocean Sciences, University of California, Santa Cruz, CA 95064, USA} \\
\aff{b}{Visiting Faculty, Scripps Institution of Oceanography} \\
\aff{c}{Department of Astronomy and Astrophysics, University of California, Santa Cruz, CA 95064, USA}  \\
\aff{d}{Simons Pivot Fellow} \\
\aff{e}{Scripps Institution of Oceanography} 
}

\abstract{
Dissolved oxygen (\doxy) is a non-conservative
tracer of interactions at the air-sea interface,
respiration and photosynthesis, and advection.
In this manuscript, we study extremes in the degree of
oxygen saturation (\so), the ratio of \doxy\ to the maximum
concentration given the water's temperature, salinity, and
depth with \so=1 critically saturated.
We perform the analysis with
the California Underwater Glider Network (\cugn), which operates
gliders on four lines that extend from the California coast
to several hundred kilometers offshore, profiling to 500\,m depth
every 3\,km.  Since $\sim 2017$, the gliders have been
equipped with a Sea-Bird 63 optode sensor to measure 
the \doxy\ content. 
We find that parcels with $\so > 1.1$, hyperoxic extrema, 
occur primarily near-shore in the upper 50\,m of the 
water column and during non-winter months.
Along Line~90 which originates in San Diego,
these hyperoxic events occur primarily in stratified waters
with shallow mixed layers. We hypothesize that photosynthesis
elevates \doxy\ in sub-surface water that can not rapidly
ventilate with the surface.
Along the three other lines, hyperoxic extrema occur almost exclusively
at the surface and are correlated
with elevated \chla\ fluorescence suggesting they are primarily
driven by blooms of photosynthesis.
We also examine hypoxic extrema, finding that 
parcels with $\so < 0.9$ and $z<50$\,m
occur most frequently along the northernmost line where
upwelling has greatest impact.
}

\begin{document}

\maketitle

%
%
%
%
%

%
\section{Introduction}
\label{sec:intro}

The dissolved oxygen (\doxy) concentration is
a fundamental property of ocean water and an
essential ingredient for life throughout the
ocean food web.  This includes the respiration
of fish and mammals as well as the growth of 
crustaceans and single-cell organisms that lie
close to the base of the food web.
\doxy\ enters the ocean through two main processes:
  (i) ventilation of air at the air-sea interface
  and (ii) biological production during processes
  like photosynthesis.
\doxy\ is removed either through respiration or if 
an increase in temperature leads to a super-saturated
state with the excess eventually fluxed through the air-sea
interface. In addition, mixing redistributes oxygen, e.g., so that
a region that is locally super-saturated 
could become less saturated.

Given the importance of \doxy, 
it has been a focus
of global analyses which assess interannual changes \citep{helm2011,schmidtko2017}. 
Best estimates for the global 
ocean indicate an $\approx 2\%$ decline in \doxy\ 
since 1960  and at a 
rate of  $\approx 10^{20} \dounit\,{\rm yr^{-1}}$ 
\citep{schmidtko2017}.
The primary causes are believed to be (1) rising
sea temperature which lowers the maximum oxygen concentration
(\oc) of water and (2) reduced ventilation to depth
owing to greater ocean stratification \citep{keeling2002}.
These processes and their impacts on \doxy\ 
have been studied in a suite of climate models 
which consistently predict a decline in \doxy\ \citep{bopp2013},
but show large variations in the regional distribution
of this signal.  Regional observations, therefore,
may impact and inform future models 
and (potentially) improve their success at predicting
future changes on a warming Earth.

At the regional level, one of the few sites with 
long-term monitoring of \doxy\ is coastal California.
Since 1984, the \calcof\ experiment has obtained \doxy\ 
measurements following standard protocol quarterly
at 66 standard stations on 6~lines running several
hundred kilometers off coastal California.
These data provided one of the first conclusive 
indications that the \doxy\ concentration is declining
\citep{bograd2008},
at least within select regions.
The origins for the decline have been proposed to be 
the advection of water with depleted \doxy\ into the
region and decreased vertical transport due to 
increased stratification near the surface.

Since 2017, the California Underwater Glider Network (\cugn)
has been measuring \doxy\ along several lines off the 
California coast 
(three since 2017 and an additional two since 2019)
with an approximately 50~times
higher profiling  frequency than \calcof.  These data have yielded a robust
estimate of the annual cycle in \doxy\  (Ren et al.\ 2004, submitted)
which resolves
seasonal trends associated with the dominant processes
along the California coast (e.g.\ upwelling). 
Furthermore, the \cugn\ dataset is now sufficiently large
(over 150,000 profiles across the four lines)
to search for rare events, i.e. outliers of 
hypoxic and/or hyperoxic water.


In this manuscript, we focus primarily on hyperoxic
events, specifically episodes with super-saturated
\doxy\ in excess of 10\%\ of the maximum oxygen concentration. 
Our principal motivation is to explore the physical
conditions that drive any such phenomena.
In turn, the results may offer insight into biological
responses in the \cugn\ associated with rare
(or common) physical events.  
Furthermore, extrema ``stress test'' models of 
complex systems which invariably include approximations
of unknown or unresolved processes.
As more advanced biogeochemical prescriptions are proposed
for regional models \citep[e.g.][]{fennel2006}, constraints related
to the incidence and generation of extrema will inform
development and reliability.
Last, hyperoxic waters have been touted as a pseudo-sanctuary
for biological life in an otherwise warming
ecosystem \citep{giomi2019}.

\section{Methods}
\label{sec:methods}

\begin{figure}[bt!] 
\centering
\includegraphics[width=\linewidth, draft=false]{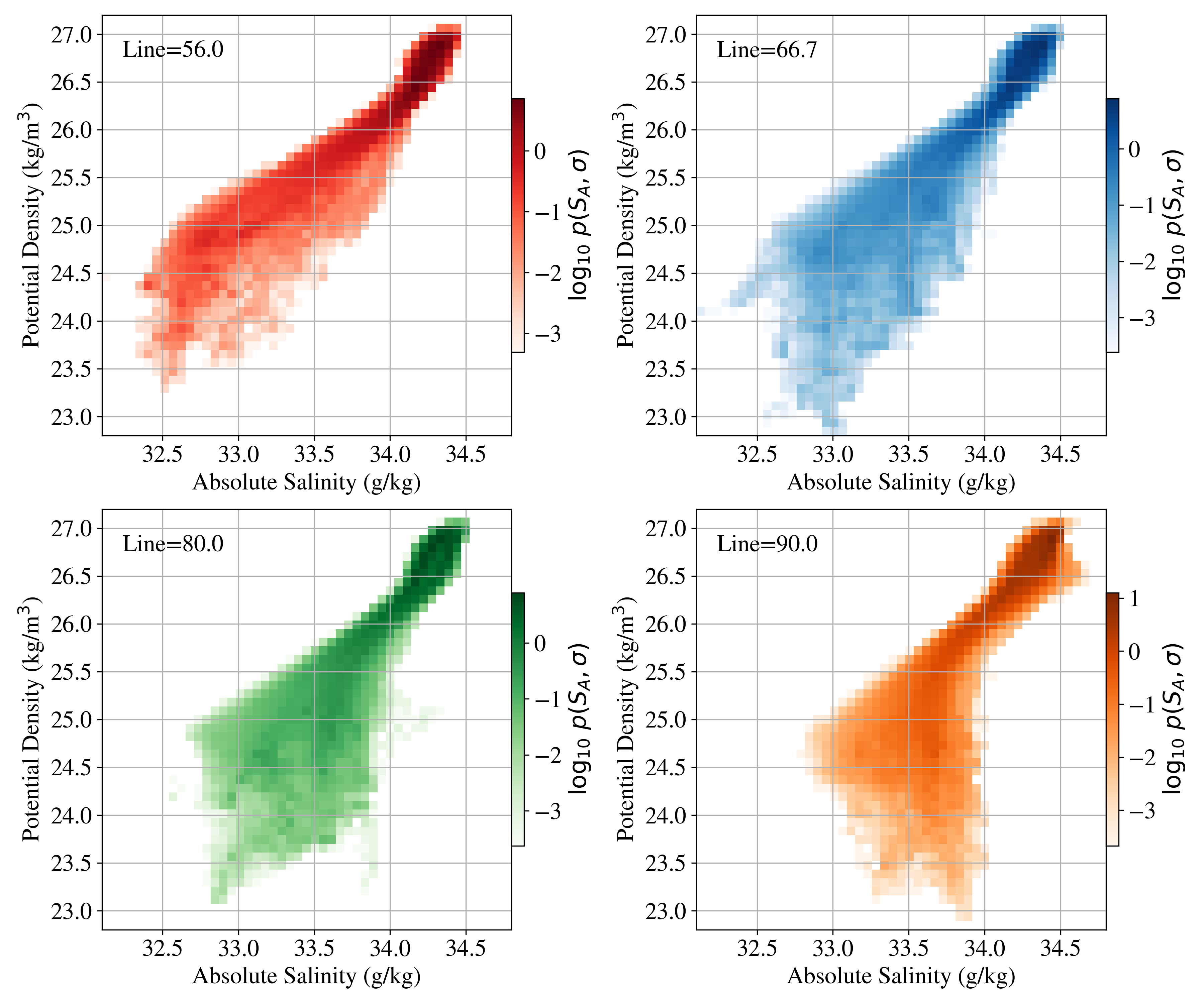}
\caption{Joint probability density functions
\pjasd\ for the conserved quantities of potential density \potd\ 
and absolute salinity \abssal\ for all water cells along
the four lines of the CUGN, restricted to those with
a valid dissolved oxygen \doxy\ measurement.  
The bins have width 
$\Delta \abssal = 0.0551 \, \salunits$ and
$\Delta \potd = 0.0898 \, \denunits$ and start
at $(\abssal, \potd) = 32.1 \, \salunits, 22.8 \, \denunits$.
On all lines, \pjasd\ peaks on the high salinity,
dense and cold water that occurs at depth ($z > 200$\,m).
The distribution then extends to lower \potd\ and \abssal\ 
where it exhibits a  wider range of values.  This 
broadening of \pjasd\ is due to several 
processes at the surface including
the injection of fresh water, evaporation, and mixing.
Note also the progression towards a higher incidence of 
lower \potd\ (warmer) and higher \abssal\ water as one
trends southward from Line~56.0 to Line~90.0. 
}
\label{fig:joint_pdfs}
\end{figure}

\begin{figure}[bt!] 
\centering
\includegraphics[width=\linewidth, draft=false]{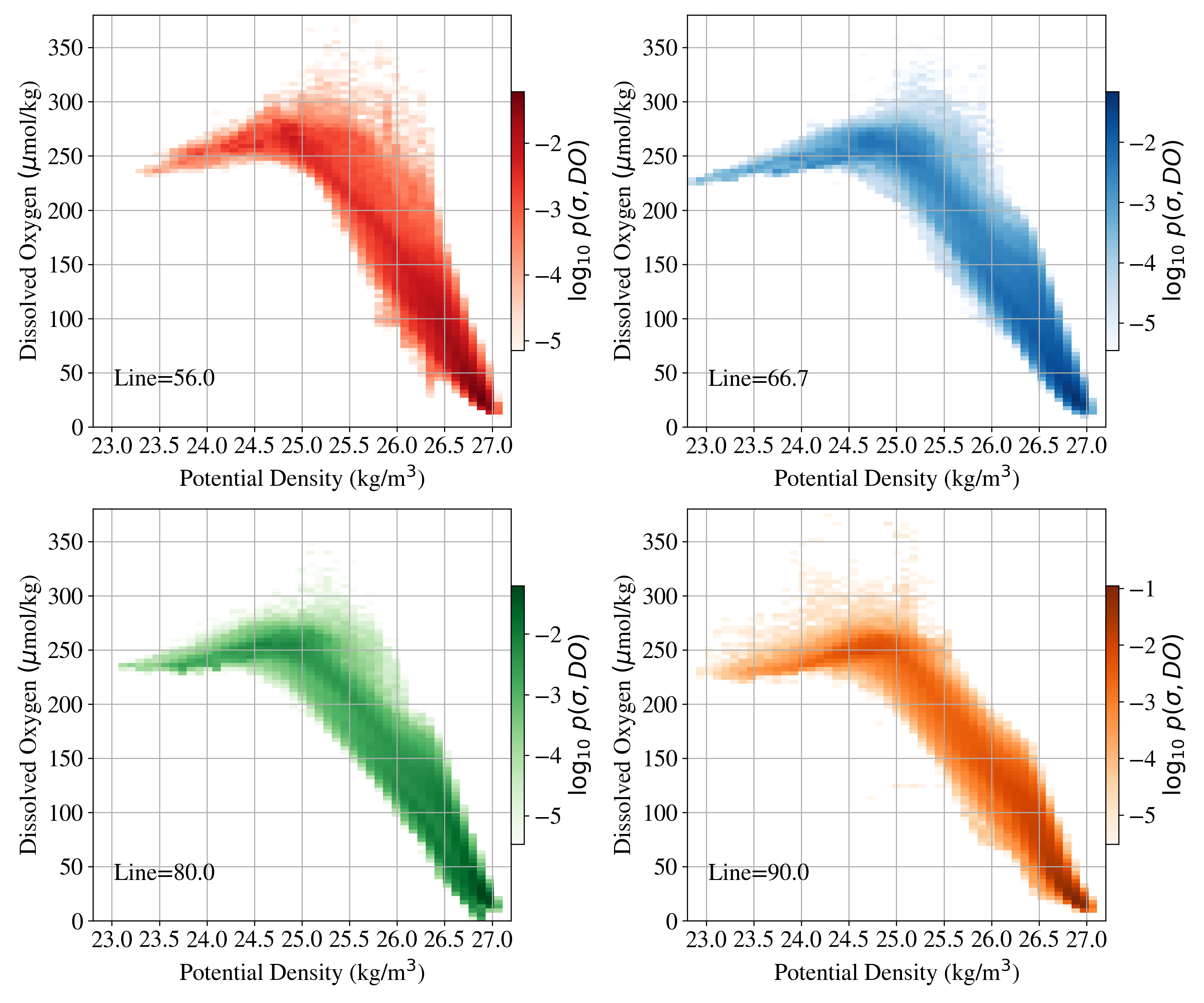}
\caption{Joint PDF for dissolved oxygen versus potential density
$p(\doxy,\sigma_0)$ for each of the four lines.
The high density water, which occur at depth, exhibits the lowest
\doxy, approaching $0 \dounit$ at the greatest depths of the 
CUGN (500\,m).  One also notes that \doxy\ 
achieves a maximum on each line at intermediate \potd.  These occur are
at or near the surface but in lower temperature water than that of
the lowest density parcels.  At these lower temperatures, the water can
maintain a higher \doxy\ concentration.
}
\label{fig:DO_joint_pdfs}
\end{figure}

\begin{figure}[bt!] 
\centering
\includegraphics[width=\linewidth, draft=false]{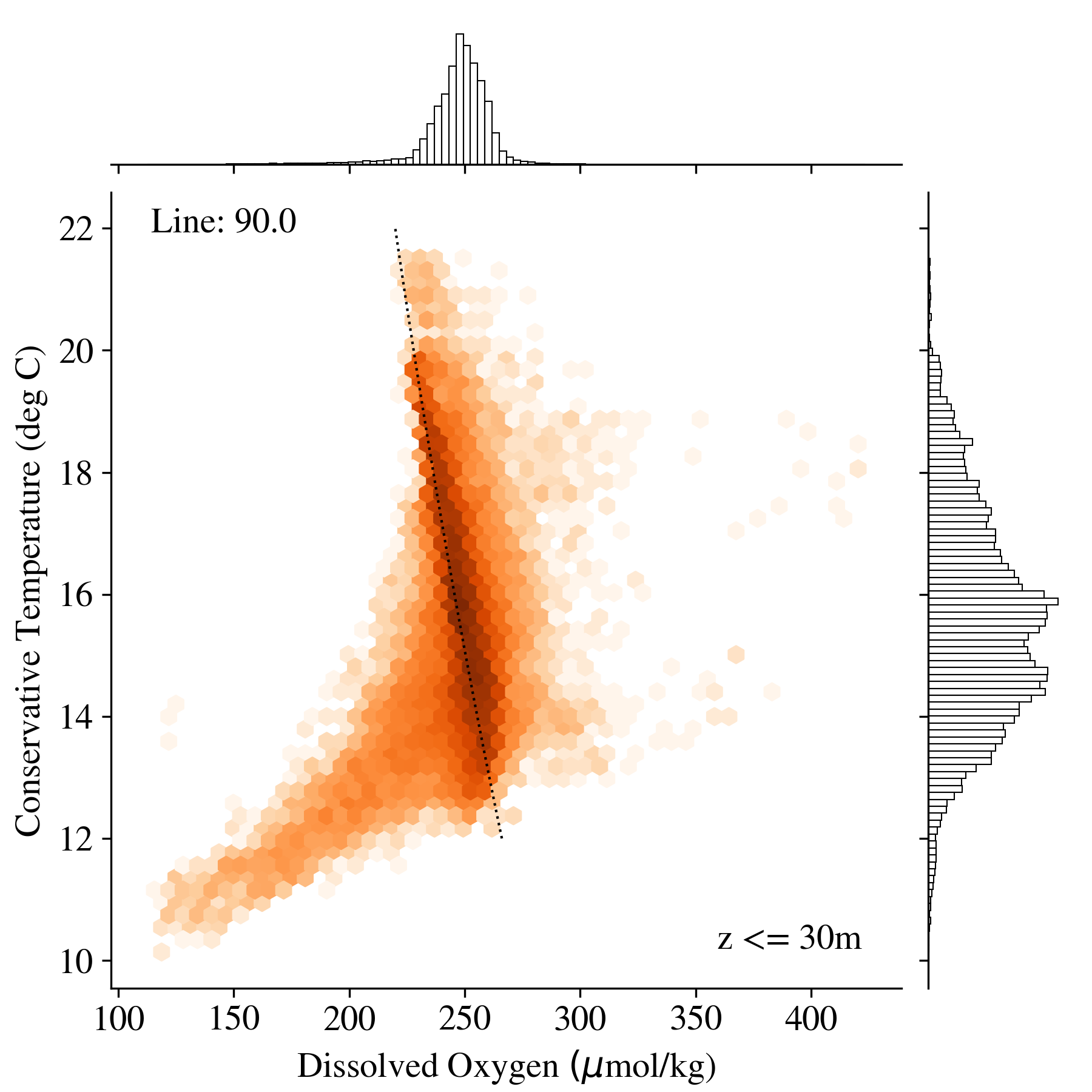}
\caption{Joint PDF \pjtdo\ of the conservative temperature
\const\ versus DO along Line~90.0 and restricted to
depths $z \le 30$\,m.
The dotted black line indicates the 
maximum oxygen concentration (OC)
for a fiducial absolute salinity
($\abssal = 33.7 \, \salunits$).
The majority of water parcels at these near-surface
depths have an oxygen saturation 
$\so = \doxy / \oc \approx 1$.
However, the figure also reveals
tails in \doxy\ to higher and lower values
indicative of under and super-saturated waters,
i.e.\ extrema that are the focus of this
manuscript.
}
\label{fig:TDO}
\end{figure}


\subsection{Data}

The California Underwater Glider Network (\cugn) uses
Spray gliders operated on 
five lines near central and southern coastal
California.  Four lines are
directed offshore (named Lines 90.0, 80.0, 66.7, 56.0 following the
\calcof\ convention) and a fifth is directed along shore (not analyzed here). 
The gliders are equipped with sensors to measure the in-situ 
temperature, salinity, fluorescence associated with chlorophyll,
horizontal velocity, and acoustic backscattering.
Since 2017\footnote{2019 for Line 56.0 when that line was added to the CUGN.}, 
the Sprays have also included a Sea-Bird~63 optode
sensor to measure the dissolved oxygen content, here reported in
units of \dounit. 
These sensors are regularly calibrated to achieve
$\approx 0.5\,\dounit$ accuracy for DO measurements in
the \cugn\ \citep{Ren2023}.

Each line is surveyed by a Spray glider 
(two on rare occasions)
traveling up to 350-500\,km offshore and back in a series of
500\,m dives, one every $\approx 3$\,km.
The sensors record measurements every 8~s or approximately
2~m horizontal distance and 1\,m in the vertical.  
In this manuscript, we analyze
the science quality, high-level data product gridded to 
cells of 10\,m  depth. 
\cite{rudnick2017} 
provides a complete description of the
optimal interpolation process used to construct the grid
and an overview of the principle properties (e.g.\ $T$, velocity)
measured along the first three lines of the \cugn\
including their annual cycles.

\subsection{Derived Quantities}
\label{sec:derived}

For each gridded dataset for each line, 
we have calculated the conserved quantities of 
absolute salinity (\abssal), conservative temperature
(\const) and potential density (\potd) following 
the Thermodynamic Equation Of Seawater - 2010 (TEOS-10) 
definitions \citep{TEOS-10}, and
adopting the grid depth and the average longitude and latitude
of the line in the calculations.  
The joint probability density function PDF of 
absolute salinity and potential density
\pjasd\ for each line
is presented in Figure~\ref{fig:joint_pdfs}.
The peak in \pjasd\ is at high salinity, high density water 
found at depth ($z> 200$\,m).
As one heads to the surface, 
the distribution extends to lower \potd\ and \abssal\ 
and exhibits a wider range of values.  
The large range of \const\ and \abssal\ at the 
surface reflect interactions that modify 
temperature (e.g.\ solar heating, cooling) and/or 
salinity (e.g.\ evaporation, rainfall, river runoff).
This study focuses on these surface waters.
Examining the four panels together, one notes the
progression towards a higher incidence of 
lower \potd\ (warmer) and higher \abssal\ water as one
travels southward from Line~56.0 to Line~90.0.
This reflects the greater influence of the California Undercurrent
which advects warmer, saltier water from the equatorial Pacific
Ocean into the CUGN \citep{rudnick2017}.

Turning our attention to \doxy, 
Figure~\ref{fig:DO_joint_pdfs} shows the joint PDFs for \doxy\ against \potd.
The high density water, which occurs at depth, exhibits the lowest
\doxy, approaching $0 \dounit$ at the highest densities and depths
of the  CUGN ($z = 500$\,m).  The PDFs also reveal that \doxy\ 
achieves a maximum on each line at an intermediate \potd.  These 
parcels are at or near the surface, but have lower temperature than that of
the lowest density parcels.  At these lower temperatures, the water can
maintain a higher \doxy\ concentration.

For each cell in the dataset we calculated the maximum oxygen 
concentration (\oc) using solubility coefficients 
derived from the data of \cite{bk84}, 
as fitted by \cite{gg92} 
and implemented in the {\sc gsw} software package \cite{gsw}. 
From \oc, one may calculate the degree of 
oxygen saturation:
$\so \equiv \doxy/\oc$ with $\so = 1$ corresponding
to a fully saturated parcel
and $\so > 1$ indicating super-saturation.

The joint PDF \pjtdo\ for \const\ and \doxy\ restricted to 
the upper 30\,m along Line~90.0 is shown in 
Figure~\ref{fig:TDO}.  
Overplotted on the figure is the $\so = 1$
contour calculated at a depth of 20\,m and for
$\abssal = 33.7 \, \salunits$.
There is a relatively sharp
ridge in the joint PDF at $\so \approx 1$ indicating
the majority of water near the surface has approximately
its maximal \oc.
At all temperatures, there is a tail of DO values 
extending beyond the $\so = 1$ contour indicating
super-saturation;  these will define \so\ extrema 
as discussed below.
One also notes water masses with $T < 13 \degc$ and
lower DO.  These under-saturated parcels from
a separate set of low \so\ extrema.

\begin{figure}[bt!] 
\centering
\includegraphics[width=\linewidth, draft=false]{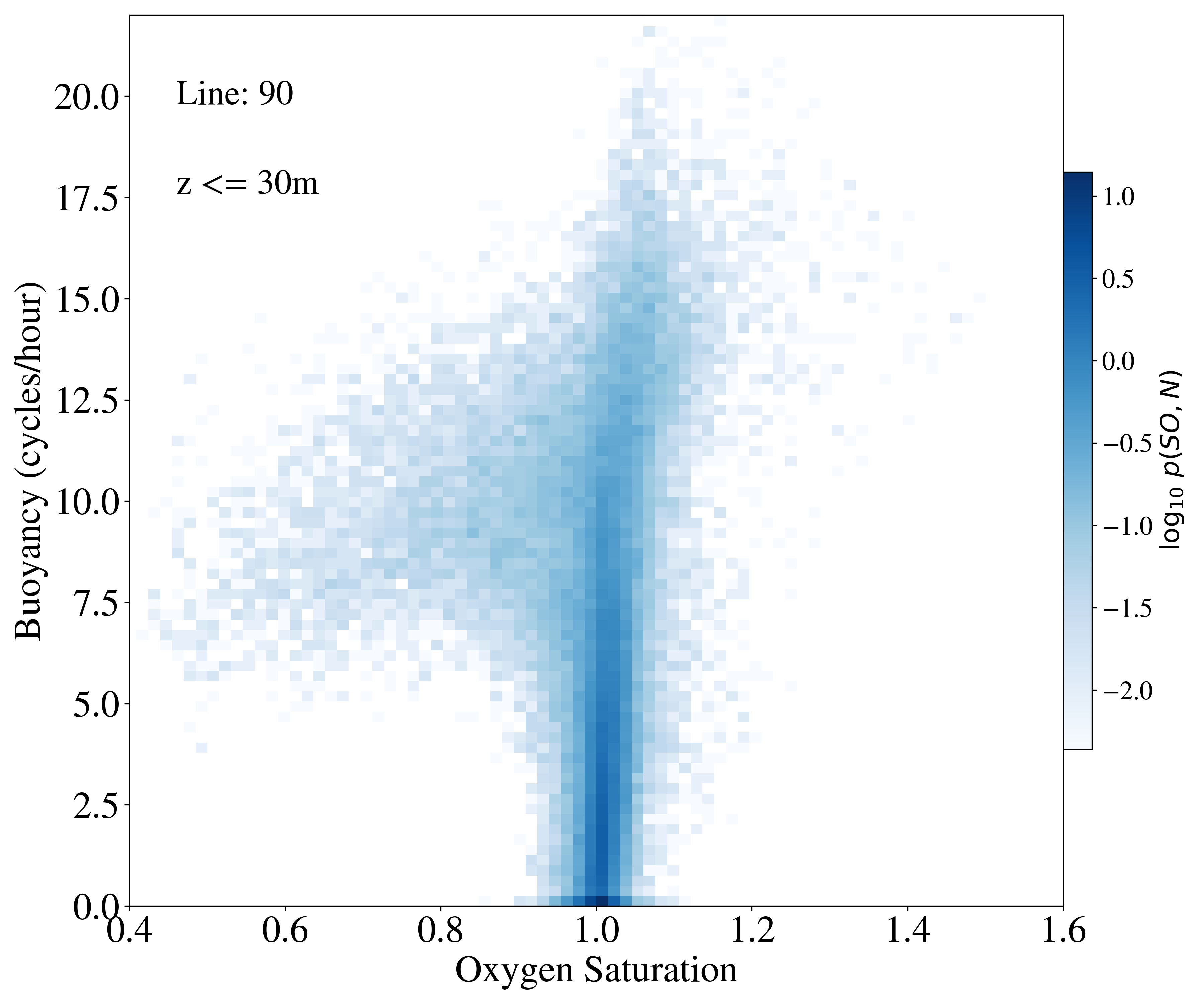}
\caption{
Joint PDF \pjson\ of buoyancy frequency (\buoy) versus
oxygen saturation (\so) for water along Line~90.0
and at depths $z \le 30$\,m.
The peak in \pjson\ occurs at $\buoy=0 \, \nunits$
and $\so = 1$, i.e. water with fully saturated
oxygen in the mixed layer.
In contrast, the extrema in \so, both 
under-saturated and super-saturated, generally have
$\buoy > 5 \, \nunits$ which suggest a 
correlation between such extrema and stratification.
}
\label{fig:SO_N}
\end{figure}

Anticipating that stratification may be influenced by
oxygen super-saturation, 
we have calculated the buoyancy frequency \buoy\ 
by taking the density gradient with depth $\buoy = (g/2 \pi \rho_0) d\potd/dz$.
Figure~\ref{fig:SO_N} shows the joint PDF for \so\ and \buoy,
\pjson\ for Line~90.0.
The peak of \pjson\ occurs at $\so \approx 1$ and $\buoy = 0$, i.e.\ 
fully saturated water with a very low buoyancy frequency.
We infer this water lies within the mixed layer
and is therefore in contact with the air-sea interface
and air-sea fluxes maintain 
\doxy\ close to its maximum oxygen concentration.
One also notes that
the tails in the \so\ distribution occur preferentially
at larger \buoy\ values, i.e.\ more stratified
columns.

\begin{figure}[bt!] 
\centering
\includegraphics[width=\linewidth, draft=false]{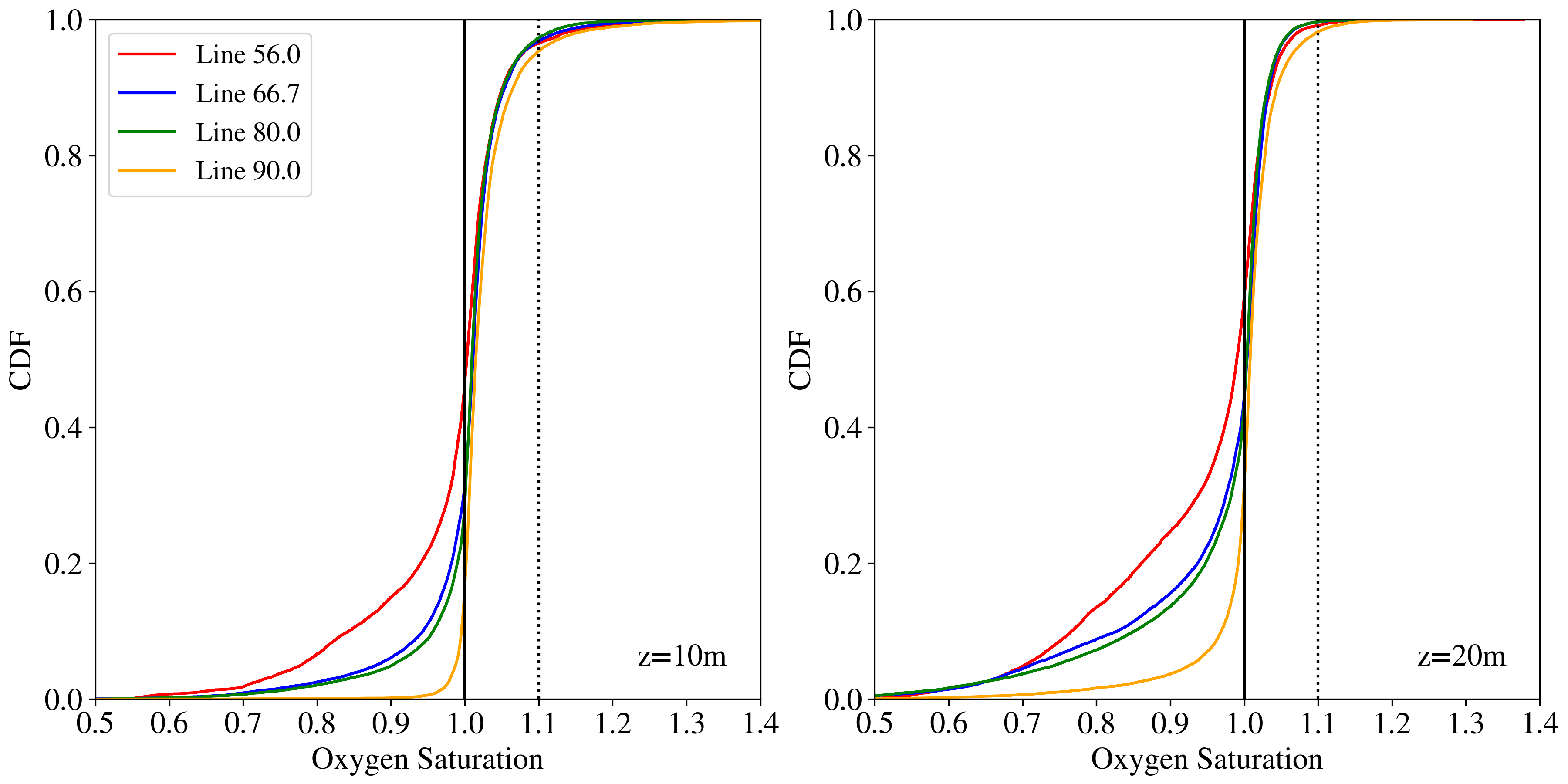}
\caption{
Cumulative distribution functions (CDFs) 
of oxygen saturation \so\ 
for all water cells in the CUGN at 
$z=10$\,m (left) and $z=20$\,m (right) depth.
We find $\so \ge 1$ for the majority of cells
at $z=10$\,m and even $\approx 50\%$ are super-saturated
at $z=20$\,m.
We have defined hyperoxic extrema as
$\so \ge \exso \equiv \valexso$ (dotted line); 
$\sim 5\%$ or fewer of the parcels have such high levels
along the lines.
One also notes a relatively high incidence of low \so,
especially along
Line~56.0; $\approx 20\%$ of its distribution 
shows $\so < 0.9$ at $z=10$\,m.
We similarly construct hypoxic extrema
defined as water with $\so \le \mnso \equiv 0.9$.
}
\label{fig:SO_CDF}
\end{figure}

\subsection{Defining Extrema in Oxygen Saturation}

To examine the incidence, nature and origins
of extrema in saturated oxygen near
the California coast, we must first define such outliers.
Figure~\ref{fig:SO_CDF} plots the cumulative distribution functions
(CDFs) of \so\ for the top 
two layers ($z=10,20$\,m) along all four lines.
It is evident from these CDFs
(and Figure~\ref{fig:SO_N})
that near the surface air-sea fluxes 
drive \doxy\ to its maximal \oc.

However, at $z=10$\,m approximately $80\%$ of the water
parcels exceed $\so = 1$ and there is a tail of
significantly super-saturated water.
Super-saturation in the ocean may be due to the active production of
\doxy\ by biological activity
(e.g\ phytoplankton growth) but may also indicate an out-of-equilibrium
condition where the water has been recently (and rapidly) heated.
On the latter point, for $\so \approx 1$ a 1\,\degc\ 
increase in the water temperature 
yields an approximately 2\%\ increase in \so:
$\Delta \so / \Delta T \approx 0.02 \, (\degc)^{-1}$.
Our scientific interests lie in searching for signatures of 
enhanced \so\ independent of the effects related
to any such rapid warming\footnote{A small signal is 
expected and observed in regions with sustained warming
\citep{emerson1987}.}. 
Therefore,  we define an upper extremum in 
\so\ (\exso) large enough to exceed changes in \so\  
from modest temperature variations (e.g.\ diurnal heating).
We believe
a conservative choice for the CUGN is $5 \degc$ corresponding to
$\exso = \valexso$. This temperature difference spans
nearly the entire range in temperature experienced
near the ocean surface throughout the analysis period
(Figure~\ref{fig:TDO}).  
While our \exso\ value is otherwise an arbitrary choice,
the results presented below are largely insensitive to
its choice for any value greater than 1.05.

Examining the CDF for \so\ (Figure~\ref{fig:SO_CDF}), 
we find
$\so > \exso$ for less than 5\%\ of the water parcels along the four lines, 
i.e.\ these are truly {\it hyperoxic extrema}.
Furthermore, the incidence of $\so > \exso$ is much less 
than 5\%\ at $z=20$\,m
where it is less than
2\%\ for Line~90.0 and $<1\%$ for the others.
Other differences between Line~90 
and the remainder will be emphasized in 
subsequent analyses.

Examining the CDFs in Figure~\ref{fig:SO_CDF} at $\so < 1$,
one also identifies tails to $\so < 0.7$ and below along each 
line at
$z=20$\,m and at $z=10$\,m (except Line~90.0).
Furthermore, we note a progression in the CDFs as one
travels from south (90.0) to north (56.0) of increasing
percentiles of low \so.  
In the following, we define {\it hypoxic extrema} of under-saturated water
as parcels with $\so < \mnso \equiv 0.9$.

\begin{figure}[bt!] 
\centering
\includegraphics[width=0.4\linewidth, draft=false]{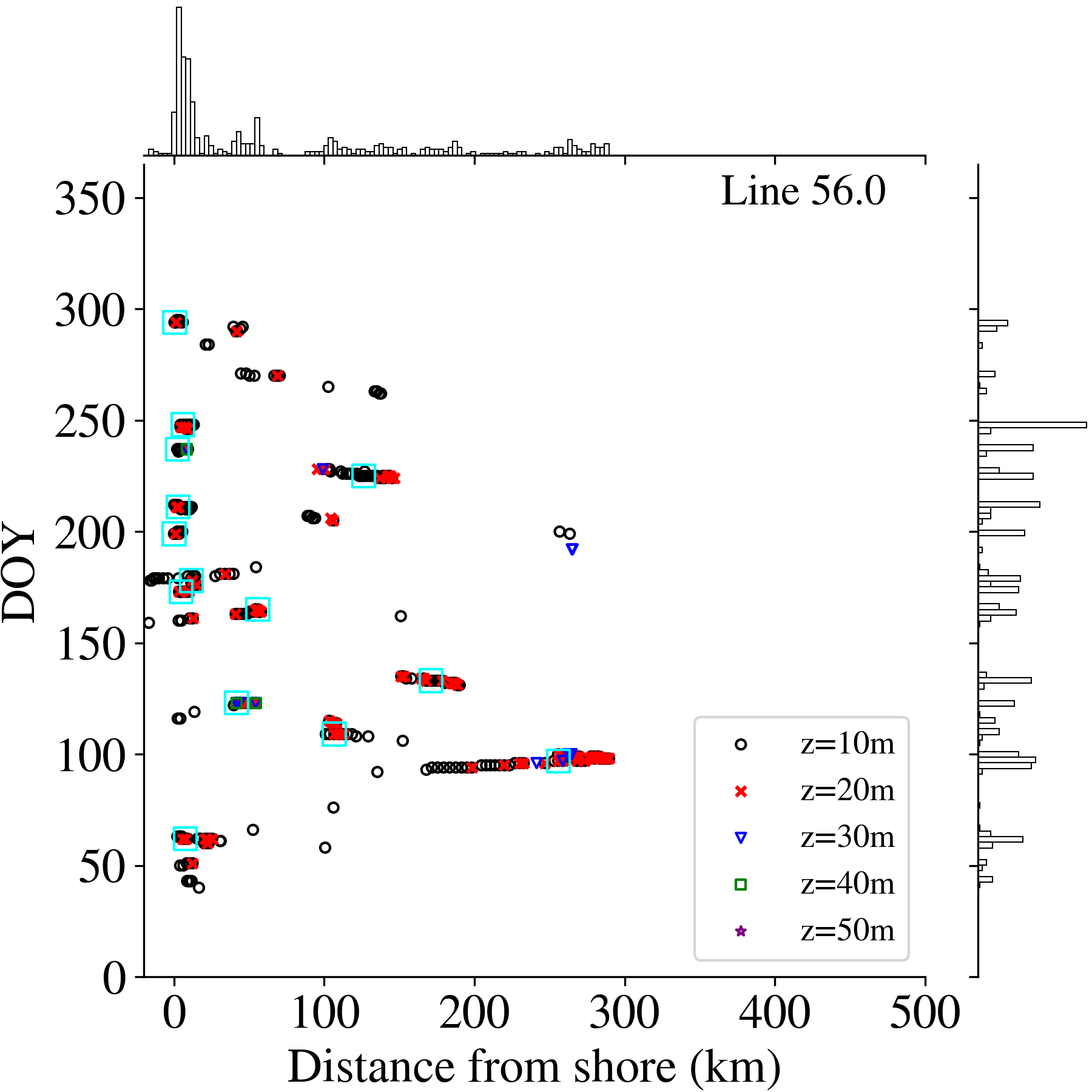}
\includegraphics[width=0.4\linewidth, draft=false]{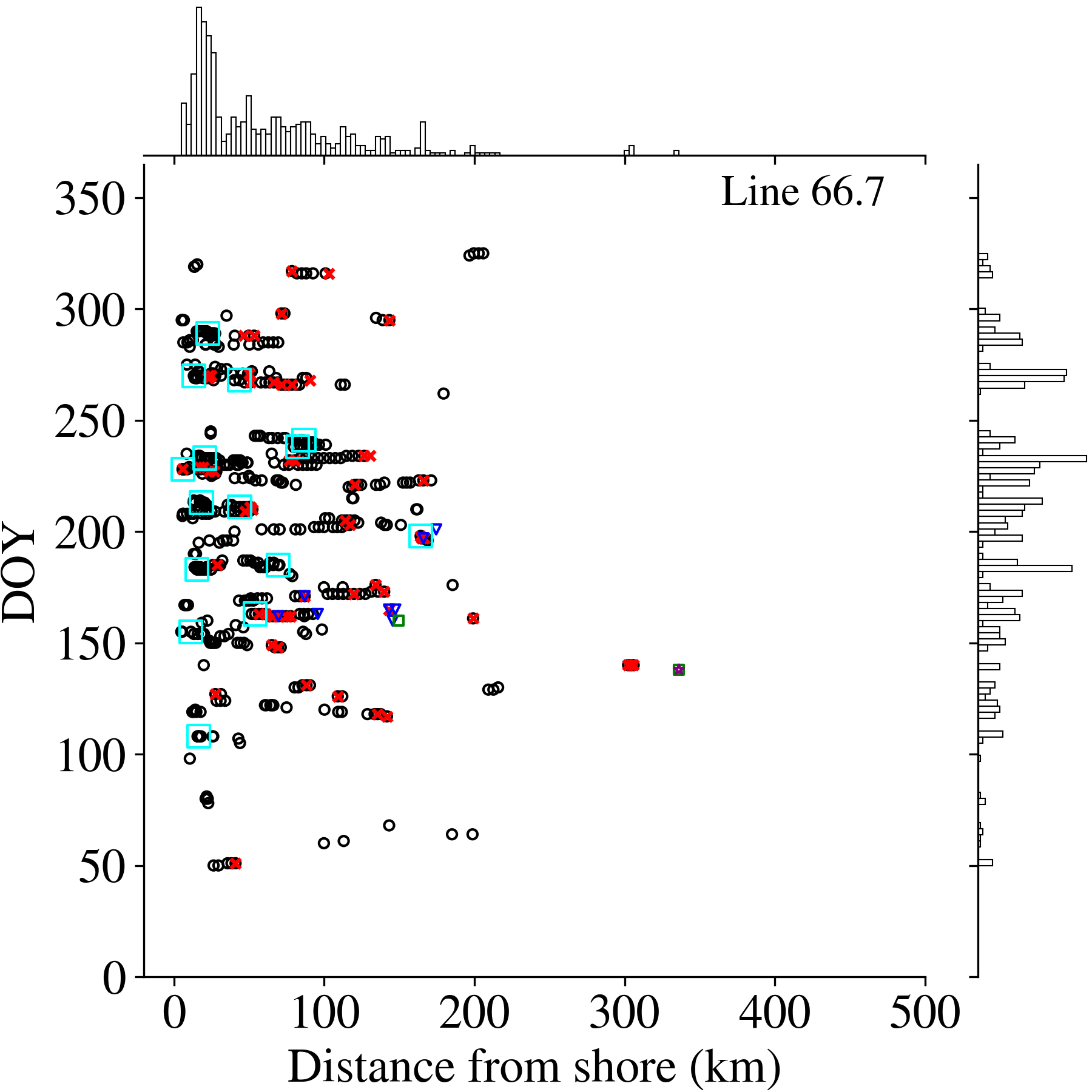}
\includegraphics[width=0.4\linewidth, draft=false]{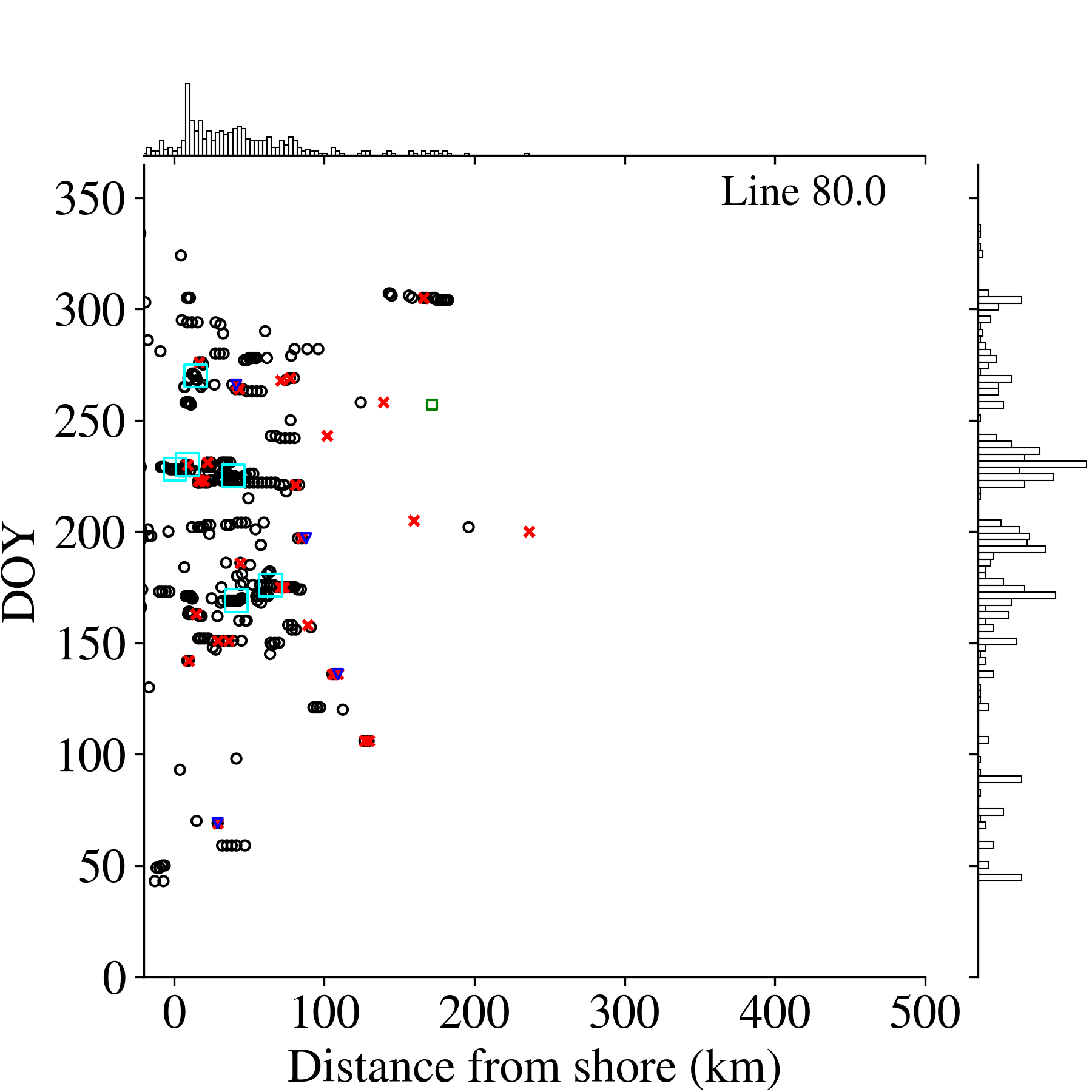}
\includegraphics[width=0.4\linewidth, draft=false]{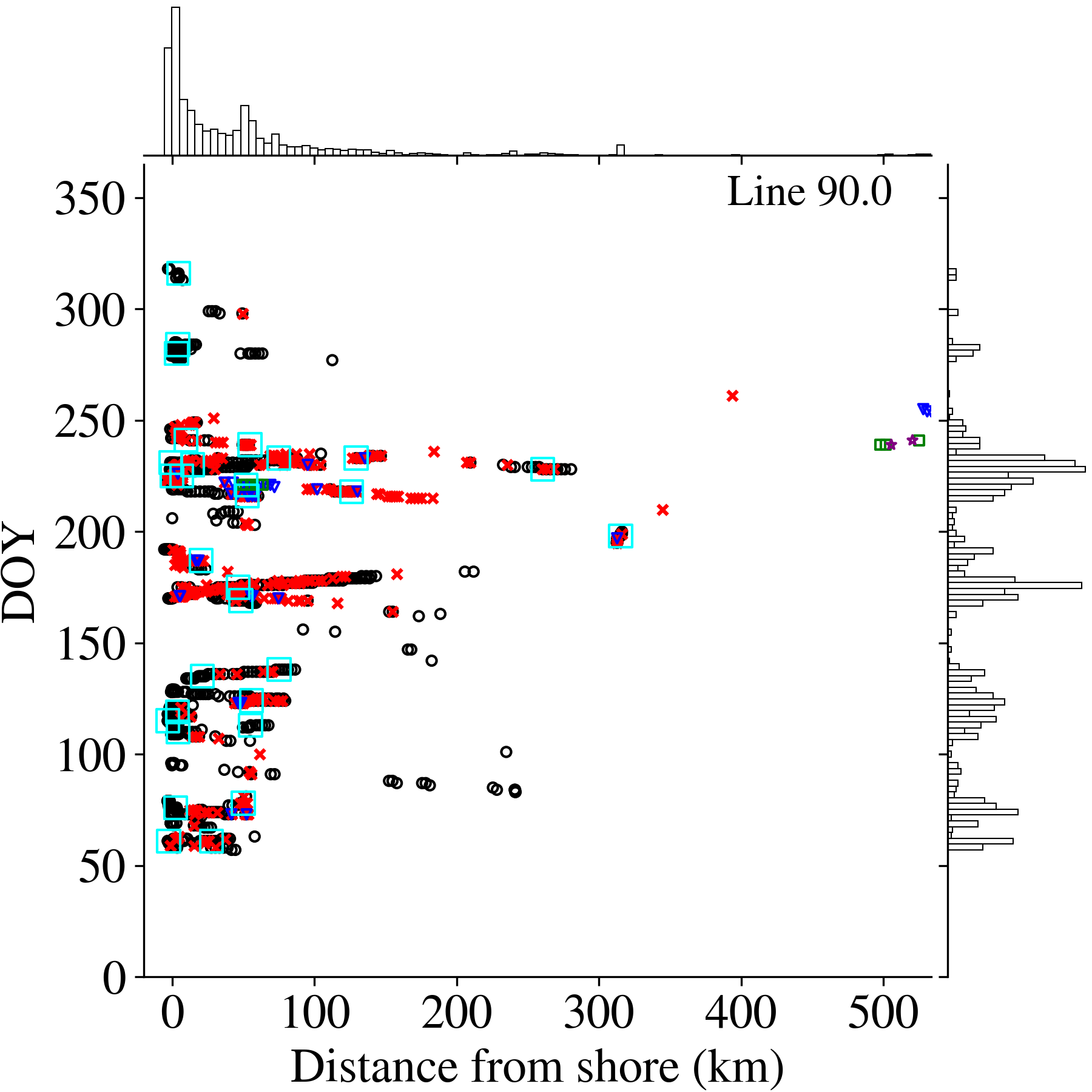}
\caption{Geographical (offshore distance) and temporal (day of year, DOY)
distributions of the super-saturated \so\ extrema along each of the four lines.
The water parcels satisfying $\so > \exso$ are marked according to their depth.
On each line, the majority occur within 100\,km of shore 
and they avoid winter months.
Only Line~90.0 exhibits a high 
incidence of hyperoxic extrema at depths $z>10$\,m.
Overlayed on the smaller symbols are large cyan
squares indicating the clustered 
events as defined in the text.
}
\label{fig:geo}
\end{figure}

\section{Results}
\label{sec:results}

\subsection{Geographical and Temporal
Distributions of \exso\ Extrema}

Consider first the geographic and temporal signatures of
super-saturated water parcels in the CUGN.
Figure~\ref{fig:geo} presents the distance from shore
and day of year (DOY) for 
hyperoxic extrema ($\so > \exso$)
on all four lines of the CUGN.
Geographically, on each line the majority 
of these extrema occur near shore, 
i.e.\ at distances less than 100\,km, and primarily
within 20\,m of the surface.
Temporally, 
the extrema avoid
Winter months (December, January, and February), 
e.g.\ on Line~90.0, 
99\%\ occur in the interval DOY=$50-300$.  
There is no significant differences in
the seasonal distribution between the lines.

Another feature apparent in the data is that the majority 
are clustered in ``\events'', i.e.\ correlated in space and time.
Quantitatively, $> 90\%$ of the extrema occur within 1~day
and 7~km of at least one other. 
With rare exceptions, the
extrema are not isolated, ephemeral,
or random excursions. 
We have generated statistically clustered
events using the Density-Based Spatial Clustering of Applications with Noise (DBSCAN) 
algorithm \citep{ester1996}
in the dimensions of depth, time, and distance offshore. 
We adopted a maximum separation of 15~days, 
10\,km, and 15\,m in depth and required a minimum of 
10~cells
to define a cluster.
We recover 64~clusters along the four lines, as marked
in Figure~\ref{fig:geo}.
From this analysis, we find an average of $\approx 2-5$ 
hyperoxic events per year on the four lines.

In a few respects, the hyperoxic extrema on 
Line~90 are qualitatively
distinct from the others.  
First, this line exhibits the highest incidence of
clustered \events\ ($\approx 5$ per year) as well
as the largest such events in cells, distance, and time.
Furthermore,  this line exhibits a higher incidence of
\so\ extrema below 10\,m and a higher fraction within
100\,km of shore.  
One anticipates that the conditions that generate \so\ extrema
north of Line~90 may be physically distinct.

Lastly, 
none of the lines exhibit an obvious interannual trend over
the 3-7~years of when \doxy\ measurements were made in the CUGN.
This baseline, however, may be too short to identify
any modest, interannual trend.

\begin{figure}[bt!] 
\centering
\includegraphics[width=\linewidth, draft=false]{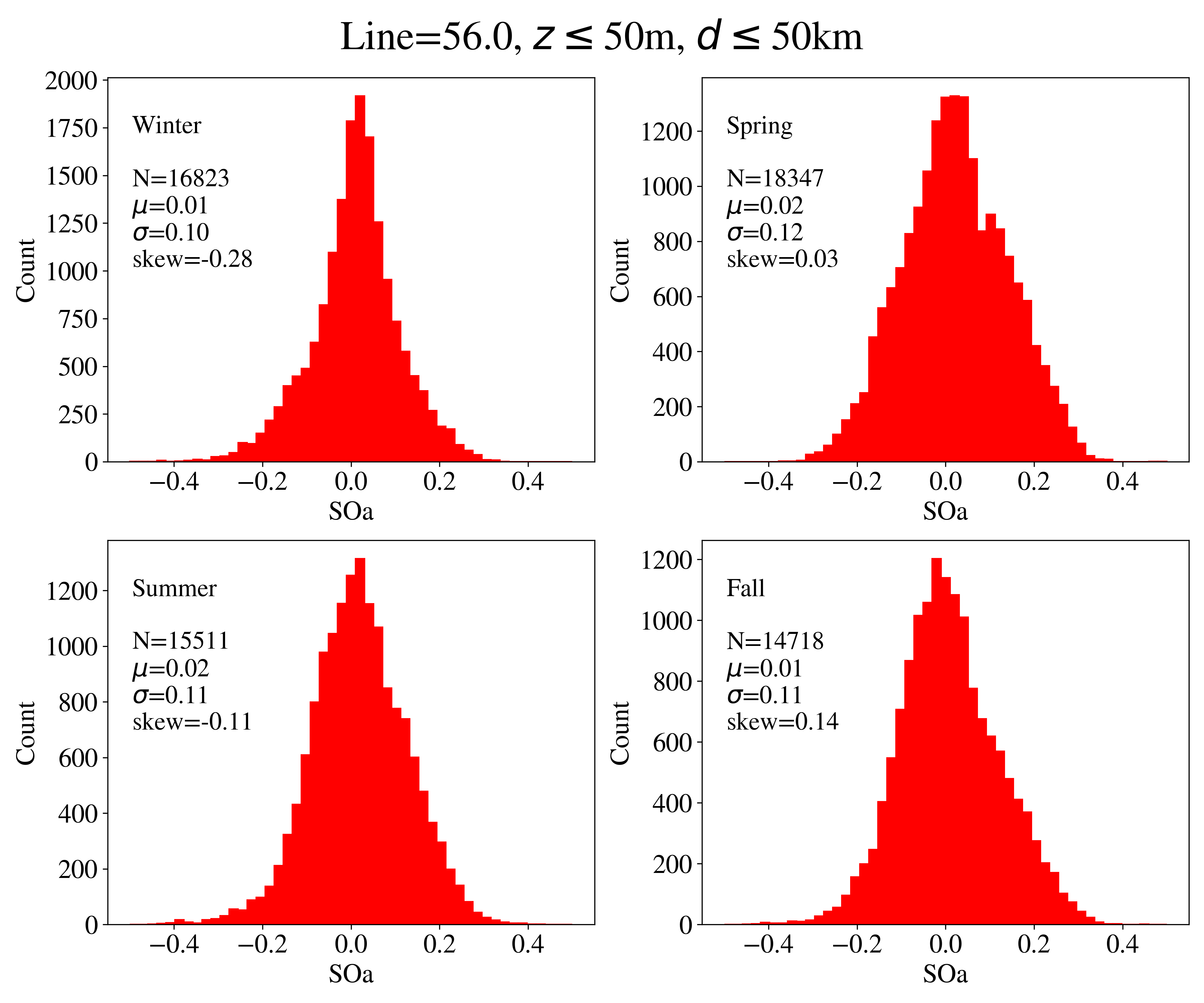}
\caption{
Anomalies of oxygen saturation relative to the annual cycle of \so, on Line~56.0
and restricted to parcels at depth $z \le 50$\,m and distance from shore
$d < 50$\,km.  The PDFs are separated by season and we provide basic
statistics for each.
}
\label{fig:SOa_56}
\end{figure}

\begin{figure}[bt!] 
\centering
\includegraphics[width=\linewidth, draft=false]{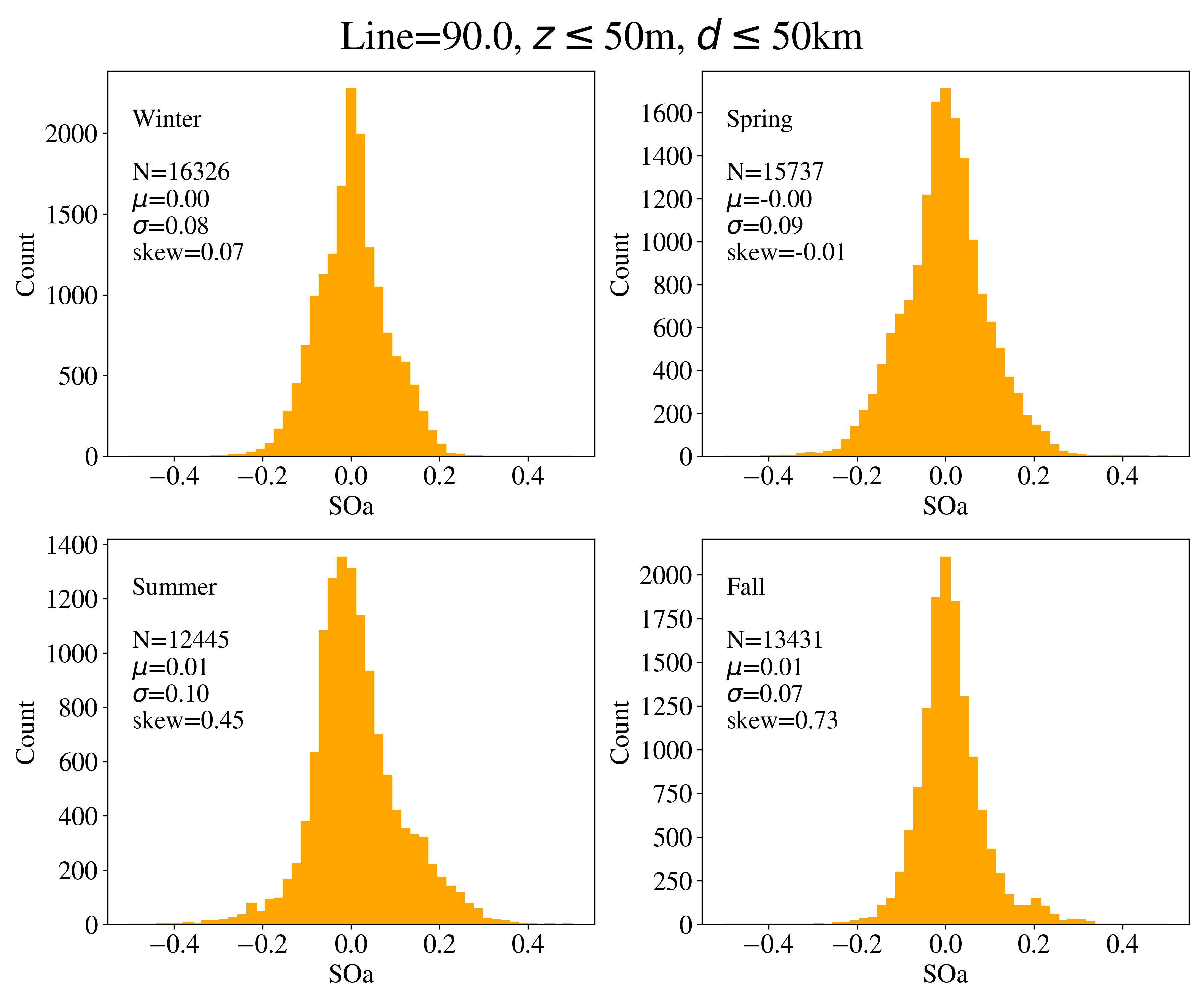}
\caption{
Same as Figure~\ref{fig:SOa_56} but for Line 90.0.
}
\label{fig:SOa_90}
\end{figure}

Figure~\ref{fig:SOa_56} and \ref{fig:SOa_90} show the seasonal variations 
in the PDFs of anomalies of \so\ defined as 
$\so_a \equiv \so - \overline{\so}$
with $\overline{\so}$ the annual cycle in \so\ (Ren et al.\ 2024, submitted)
using the method described in \cite{rudnick2017}.
On Line~56.0, the distribution shows a significant, negative skew in the winter
months indicating a higher incidence of under-saturated waters, perhaps related
to upwelling activity.
Later in the year, the PDFs show a small, positive skew, i.e.\ higher $\so_a$
values.
Contrast these results with the PDFs from Line~90.0 which never show a negative
skew and which show large positive skews in the Summer and Fall months.
These skews are driven by the hyperoxic extrema.

\subsection{Representative Examples of $\so > \exso$ Extrema}
\label{sec:exso_examples}

\begin{figure}[bt!] 
\centering
\includegraphics[width=0.95\linewidth, draft=false]{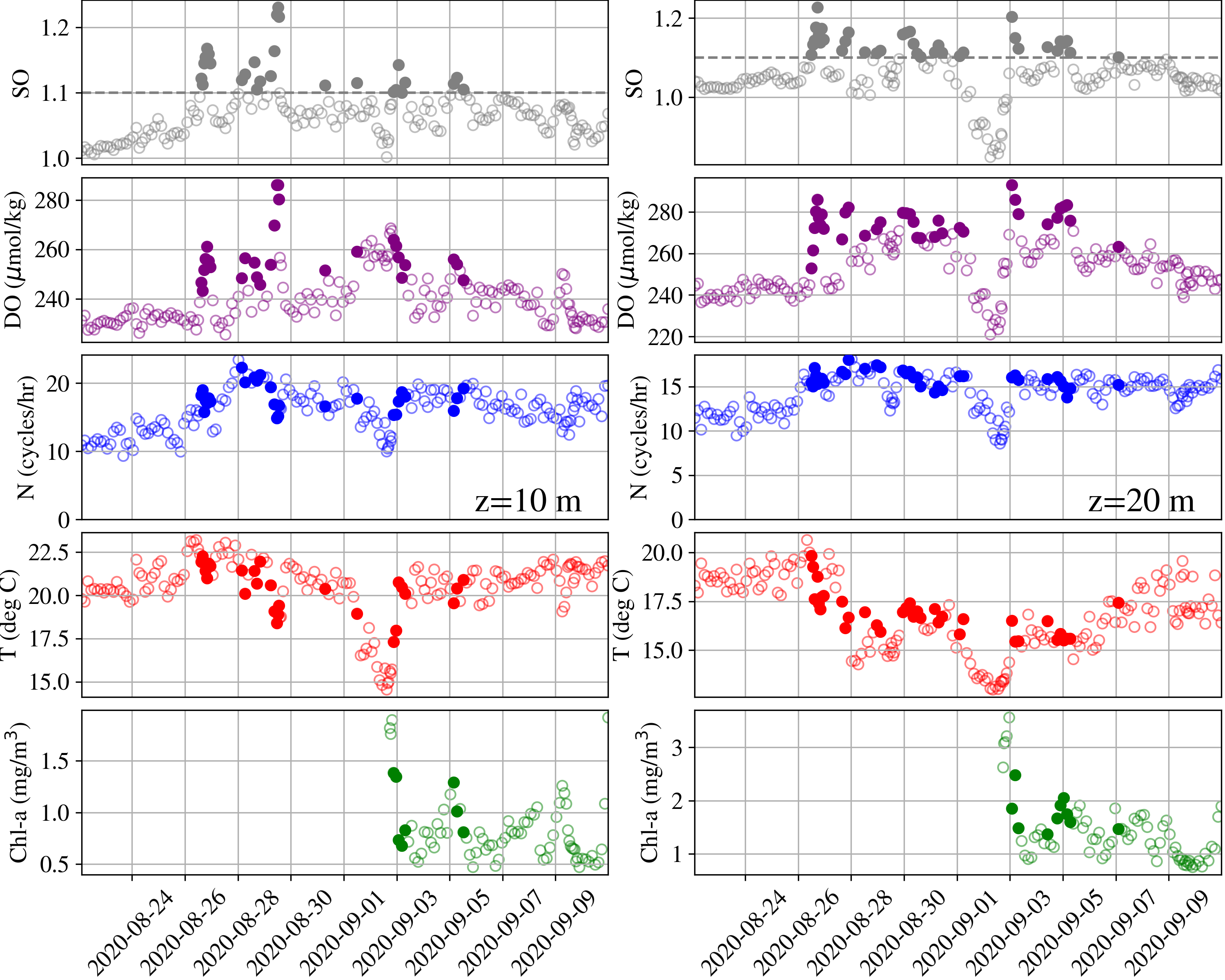}
\caption{
\so, \doxy, \buoy, $T$, and \chla\ measurements
for an approximately 3~week interval 
spanning an hyperoxic event on Line~90.0
at depth $z=10$\,m (left) and $z=20$\,m (right).
On the first/last day of the interval, the glider
is 132/86~km offshore and turns around on 2020-09-02.
Filled symbols highlight the $\so > \exso$ extrema,
which for this event are primarily at $z=20\,$m.
}
\label{fig:ex_90A}
\end{figure}

We now examine several representative \events\ of $\so > \exso$ extrema
within the CUGN.
Figure~\ref{fig:ex_90A} shows the \so, \doxy, $T$, \chla, and \buoy\ measurements
for the water parcels near-shore on Line~90.0 in late-August 
to mid-September 2020.
For this event, the majority of extrema occur sub-surface
at $z \approx 20$\,m.
Not surprisingly 
the extrema show very high DO, generally
at $\approx 270-290 \, \dounit$, and the \so\ values are high
relative to the other parcels with $\so < \exso$.
Furthermore, nearly all of the extrema have an elevated
buoyancy frequency of  $\buoy > 15 \, \nunits$. 
As regards temperature, the values for the 
extrema tend towards the median for
the water probed during this 3~week interval. 
At both $z=10$\,m and $z=20$\,m, the warmest waters 
probed rarely 
satisfy the \exso\ threshold, despite the fact that higher
temperature means a lower, maximum oxygen concentration
(i.e.\ higher SO for the same DO).
Lastly, the \chla\ values of the extrema generally exceed
$1 \chlunits$ but do not track the highest values
recorded during the interval.

\begin{figure}[bt!] 
\centering
\includegraphics[width=0.95\linewidth, draft=false]{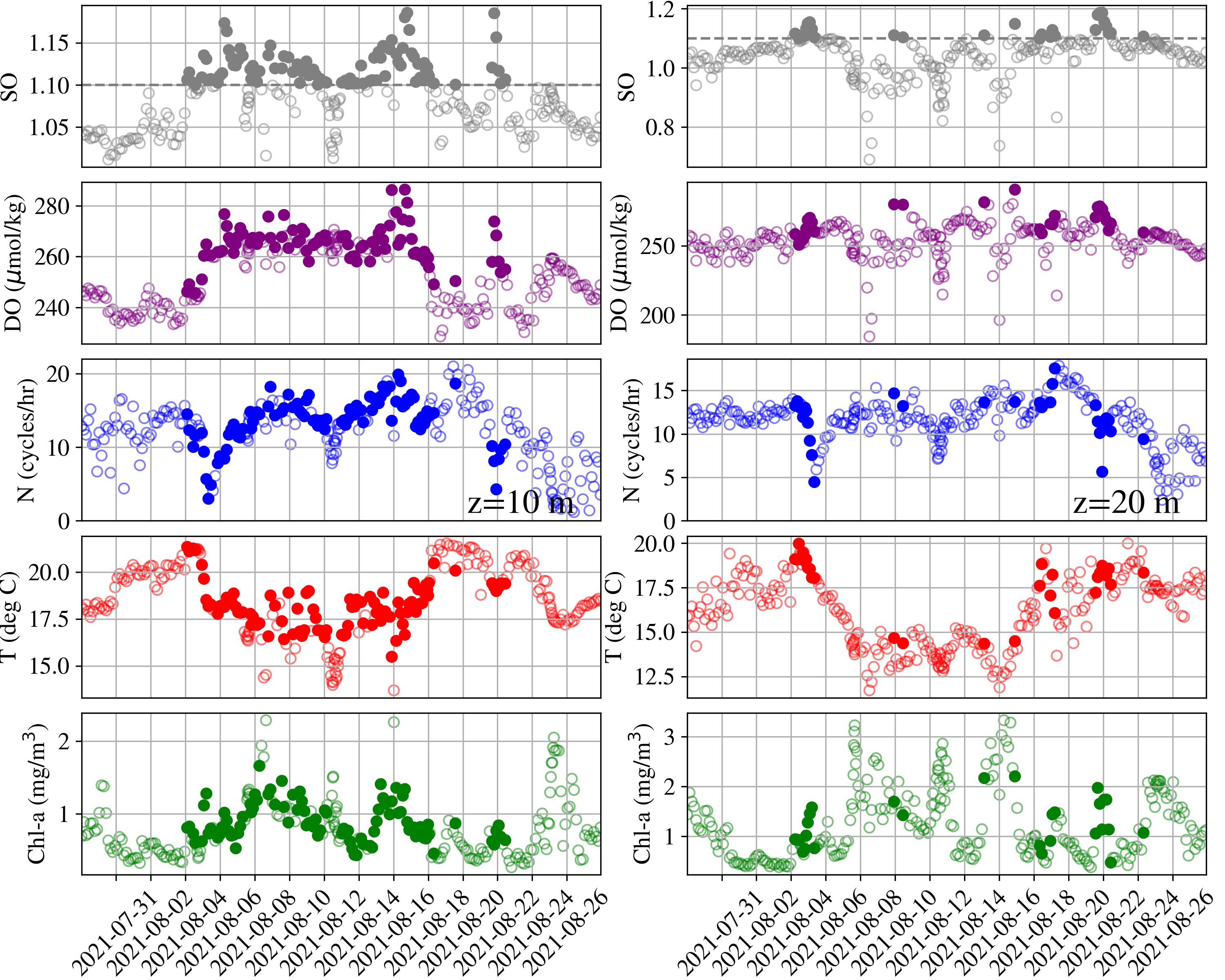}
\caption{As in Figure~\ref{fig:ex_90A},
a representative example of a 
super-saturated event along Line~90.0 but now
primarily within the upper, $z=10$\,m layer.
We otherwise find the hyperoxic extrema have similar
characteristics (high \doxy, high \buoy\ and
intermediate temperature and \chla).  
}
\label{fig:ex_90B}
\end{figure}

Figure~\ref{fig:ex_90B} shows another near-shore \event\ 
along Line~90.0, in this case during August 2021.
Here, the extrema are primarily in the upper layer
($z = 10$\,m).
The hyperoxic extrema of this event exhibit similar characteristics
as for the first example: 
  high \doxy\ values ($>260 \dounit$),
  elevated $N$ ($>10 \nunits$),
  and intermediate temperature and \chla\ concentration.
Also similar to the first example,
the warmer waters show systematically lower \doxy\ 
and therefore do not satisfy $\so > \exso$,
although they are generally super-saturated.

\begin{figure}[bt!] 
\centering
\includegraphics[width=0.95\linewidth, draft=false]{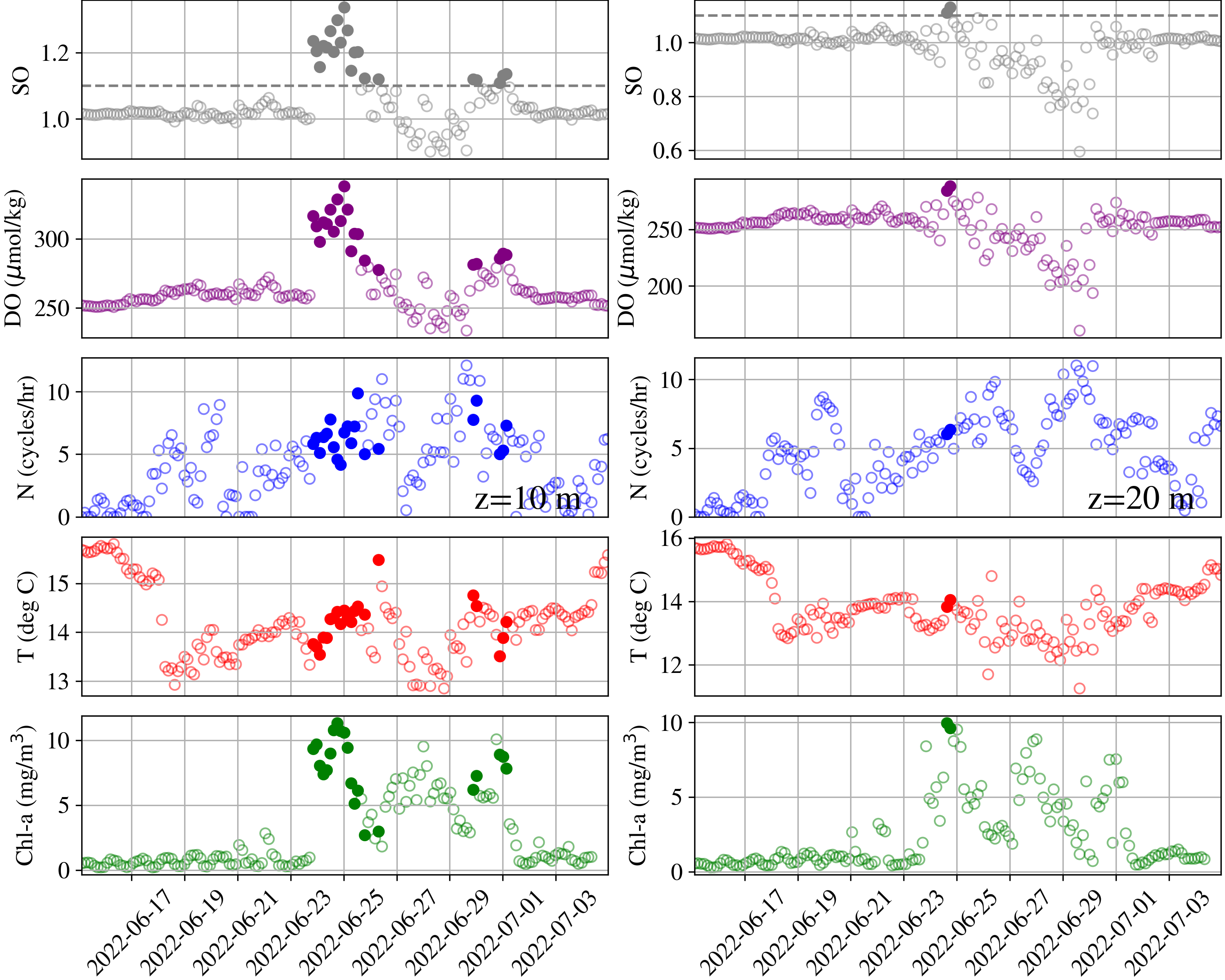}
\caption{Example of a hyperoxic \event\ along Line~80.0.
In contrast to the examples from Line~90.0, 
the \events\ along the other lines tend to have
shorter duration, lower \buoy\ values, and much
higher \chla\ concentration.
}
\label{fig:ex_80}
\end{figure}

Now consider a representative example (Figure~\ref{fig:ex_80}) 
of an $\so > \exso$ event
from one of the other CUGN lines. 
Similar to the \events\ from Line~90.0, the super-saturated
water shows very high \doxy\ values (here, 
primarily above 300\,\dounit) and non-anomalous 
temperatures.
In contrast to Line~90.0, however, the extrema
in this \event\ are characterized by
lower buoyancy frequency and 
extreme \chla\ concentrations.

\begin{figure}[bt!] 
\centering
\includegraphics[width=0.95\linewidth, draft=false]{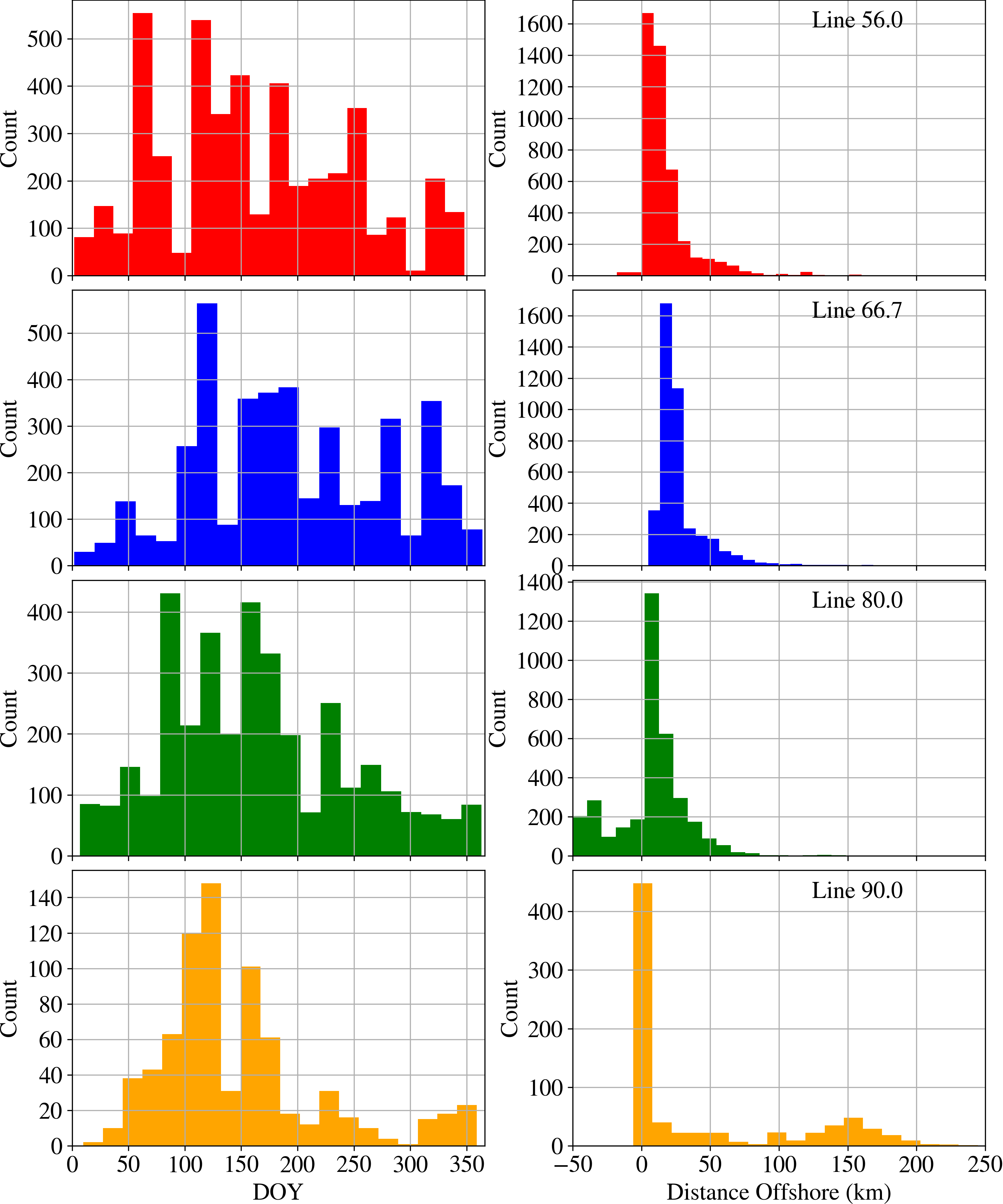}
\caption{Temporal (DOY, left) and geographical (distance from shore, right)
distributions of the hypoxic extrema ($\so < \mnso$) near the surface
($z \le 20$\,m) along the four lines.
These extrema are located primarily near shore, in regions most affected by upwelling.
For Lines 80.0, 90.0, they also occur most frequently in spring and early summer.
}
\label{fig:geo_low}
\end{figure}

\subsection{Representative Examples of $\so < \mnso$ Extrema}

Figure~\ref{fig:geo_low} describes the geographic
and temporal distribution of under-saturated,
near-surface ($z \le 20$\,m) waters along the 
four lines.  The overwhelming majority of these
extrema are located near-shore;
except for Line~90.0, $\approx 99\%$ lie
within 100\,km.
In addition, the majority occur preferntially
within spring or early summer (DOY=50-200).

\section{Discussion}
\label{sec:discuss}

The previous section described the basic properties
and distributions
of \so\ extrema and examined several representative
examples.  We turn now to discuss plausible origins
and comment on the importance of this work for
studies on coastal California
and the broader ocean.

\subsection{The Principal Drivers of Hyperoxic Extrema along Coastal California}

In the CUGN, hyperoxic \events\ occur most 
frequently within 
100\,km of shore, within 30\,m of the ocean surface, and 
in non-winter months.  They frequently occur in \events\ lasting
several days to several weeks.

\begin{figure}[bt!] 
\centering
\includegraphics[width=0.95\linewidth, draft=false]{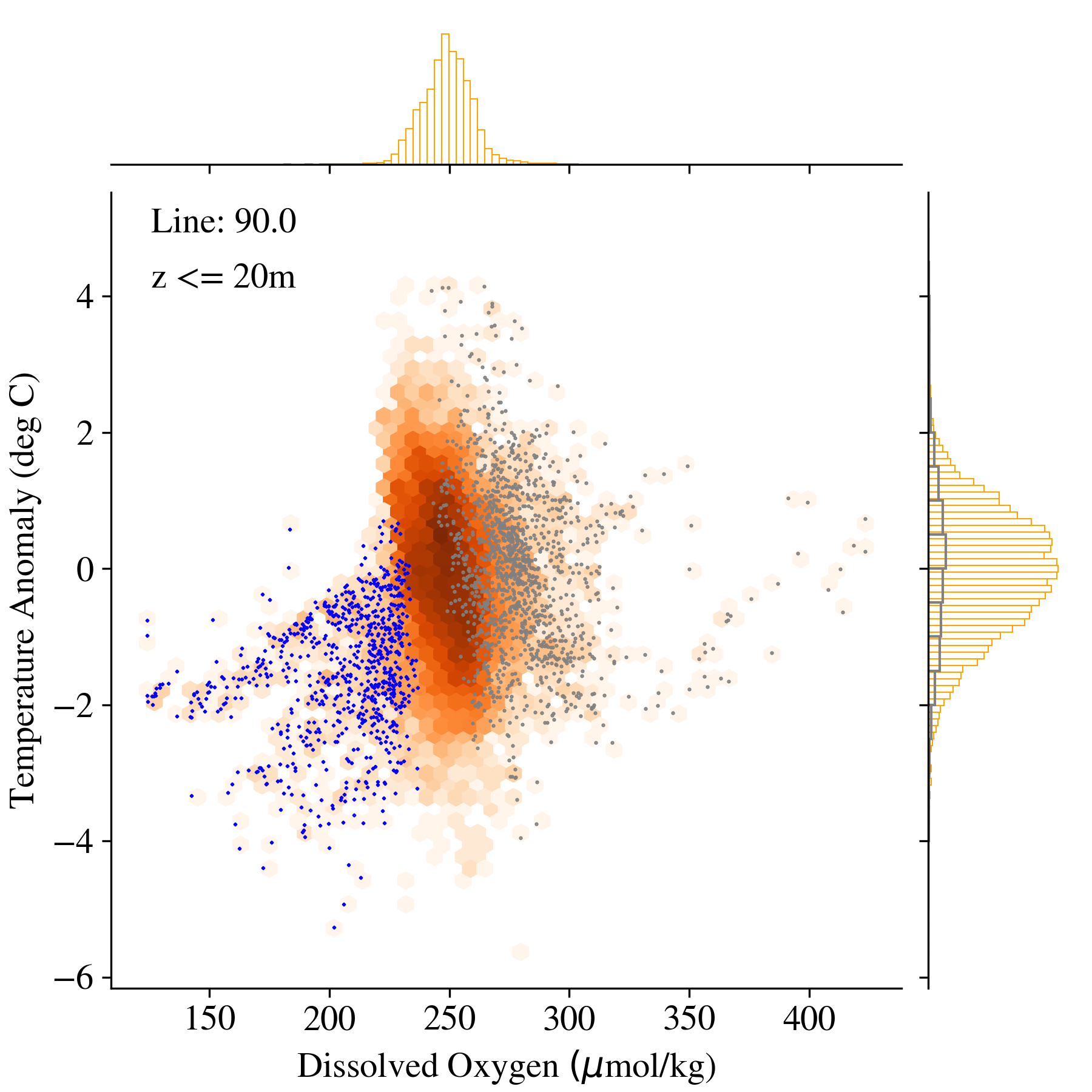}
\caption{Joint distribution of temperature anomalies
measured relative to the annual cycle versus dissolved
oxygen restricted to Line~90.0 and $z \le 20$\,m.
Overplotted as dots
on the joint PDF are the hyperoxic (gray)
and hypoxic extrema (blue).
The former are distributed uniformly in temperature
(compare the histograms) whereas the hypoxic extrema
occur almost exclusively at colder than average
temperature.
}
\label{fig:Tanom_90}
\end{figure}

As we explore the primary drivers of such extrema,
we remind the reader that for the DO in a
given water parcel to (at least temporarily) exceed its 
maximum oxygen concentration,
either the \oc\ has been lowered by increasing the temperature
(without lowering DO) and/or biological productivity 
has increased DO within the water.
In the representative examples along Line~90.0
(Figures~\ref{fig:ex_90A}, \ref{fig:ex_90B}), 
there is no signature of elevated temperature in the 
hyperoxic extrema.  
In fact, if anything the $\so > \exso$ extrema 
show cooler temperatures than their surrounding waters.
To further examine the potential role of temperature, 
for Line~90.0 we have calculated the 
temperature  anomaly of the extrema relative 
to the annual cycle calculated
by \citep{rudnick2017}.  
Figure~\ref{fig:Tanom_90} compares these with all 
parcels on Line~90.0 restricted to $z \le 20$\,m. 
The distributions of both the 
hyperoxic extrema and the control sample
peak at 0\,deg\,C with a symmetric distribution spanning 
$\approx \pm 2\,\degc$.  
There is no systematic trend of the $\so > \exso$
extrema towards unusually warm temperatures, and nothing
approaching  the $\approx 5\,\degc$ that would
be required to bring marginally saturated water to
$\so > 1.1$.
We conclude, therefore, that the primary driver of 
extreme super-saturated water is not a rapid rise in temperature. 

This requires, therefore, that the extrema originate
from an elevated \doxy\ concentration.
Indeed, when examining the representative examples 
we stressed the presence of 
high \doxy\ values both in absolute terms
and relative to nearby water parcels not satisfying
the \exso\ threshold.
We test this inference statistically by
comparing the \doxy\ distribution of the 
$\so > \exso$ extrema along each line
to that from a custom control sample.
For the latter, we used the joint PDF 
grid of absolute salinity and potential density
shown in Figure~\ref{fig:joint_pdfs} 
restricted to those cells with at least
50~samples. 
For every hyperoxic extremum, we take at random 
5~parcels in the same \abssal,\potd\ bin and the
ensemble is the matched control sample.

\begin{figure}[bt!] 
\centering
\includegraphics[width=0.95\linewidth, draft=false]{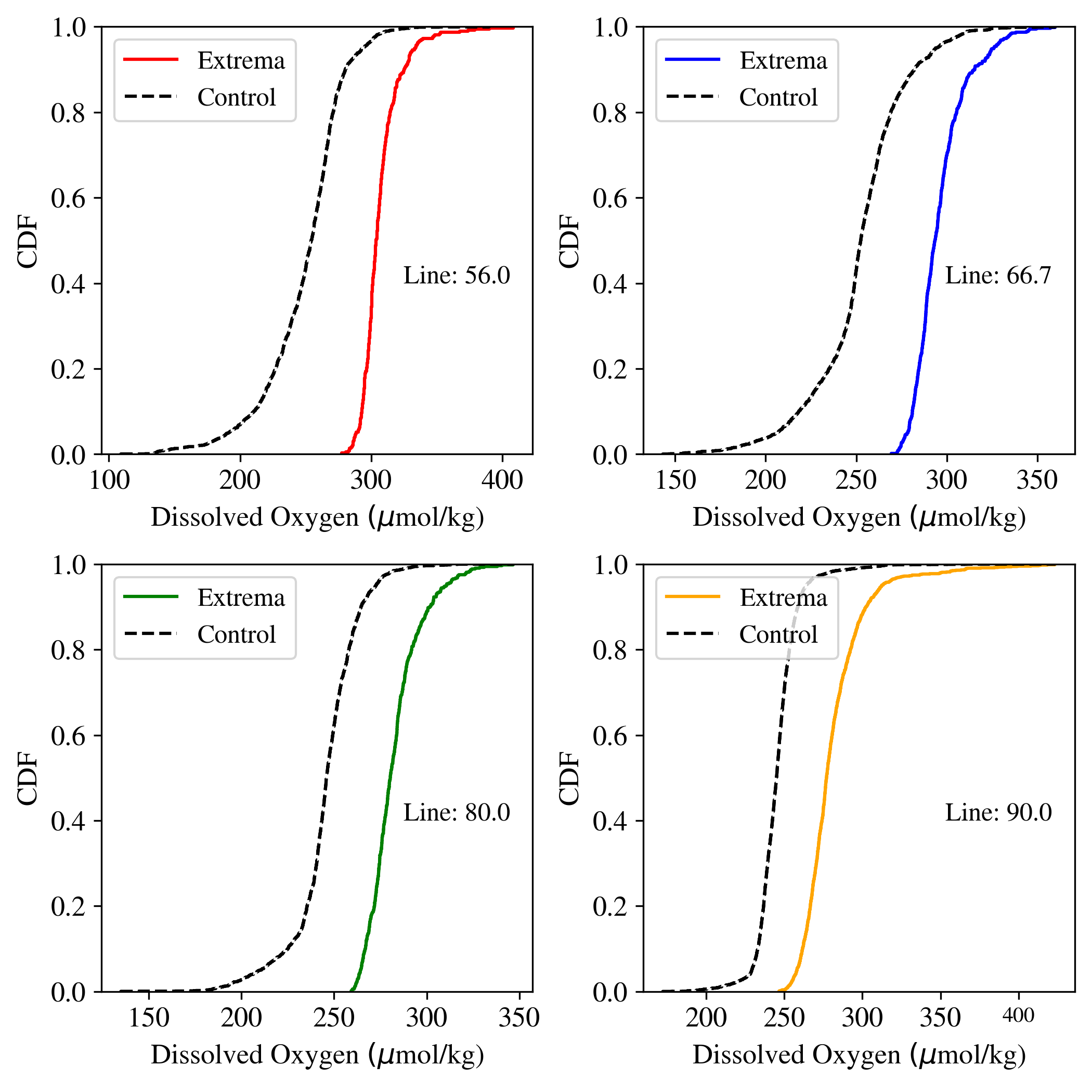}
\caption{Cumulative distribution functions
(CDFs) for dissolved oxygen for the $\so > \exso$
extrema (solid lines) along each \cugn\ line.
For each line, we have generated a control sample
matched to the extrema in absolute salinity and
potential density and restricted to depths
$z \le 50$\,m.  The CDFs of the extrema show
systematically higher \doxy\ values, as expected.
}
\label{fig:DO_cdfs}
\end{figure}

Figure~\ref{fig:DO_cdfs} shows the CDFs of
the \doxy\ distributions for the extrema versus
that of the control sample.
Clearly, the \doxy\ distribution of 
the extrema well exceed that of the control sample confirming
a systematic \doxy\ excess.
We now consider conditions that may 
drive this excess.

\begin{figure}[bt!] 
\centering
\includegraphics[width=0.95\linewidth, draft=false]{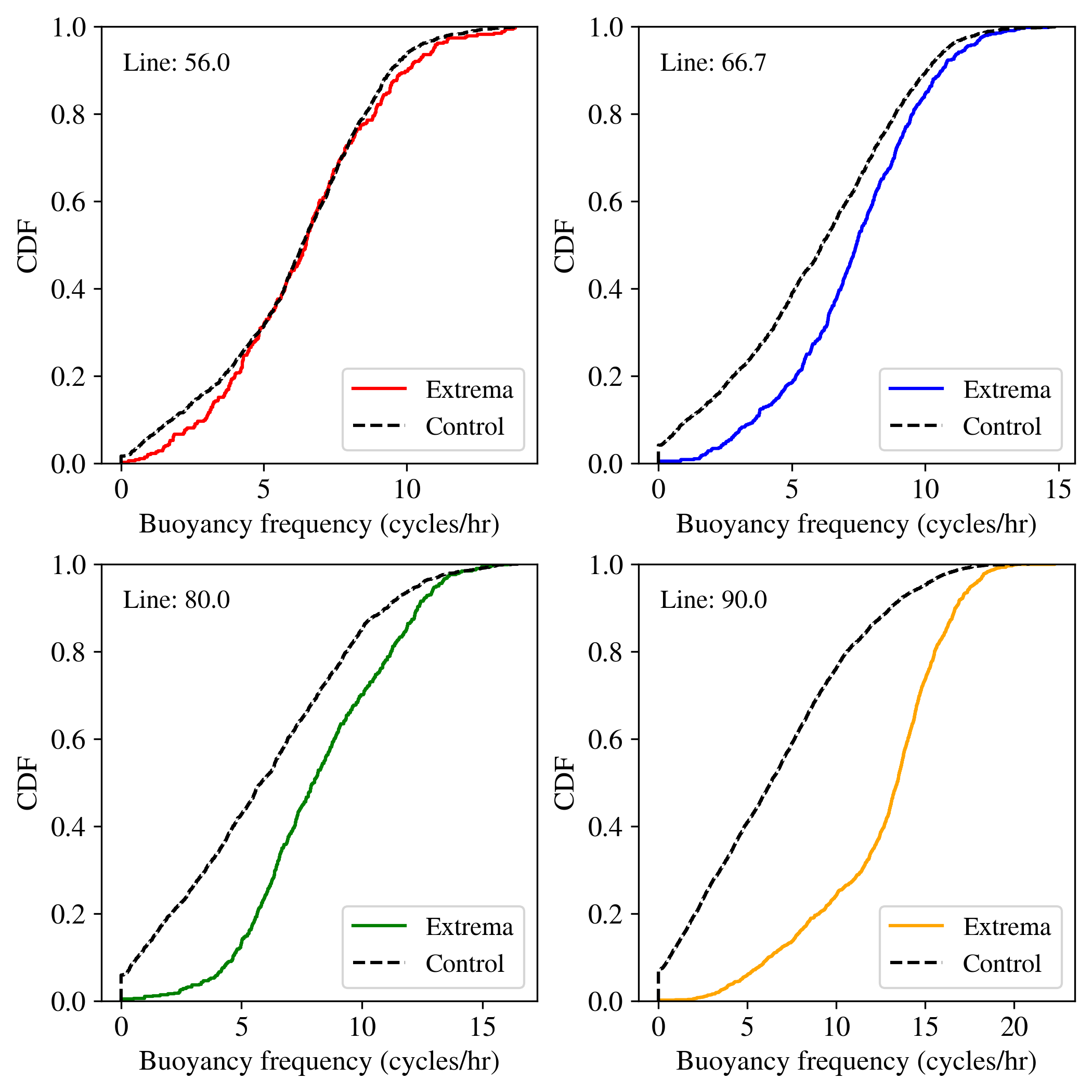}
\caption{Similar to Figure~\ref{fig:DO_cdfs}
but for buoyancy frequency.
For these results, the extrema show increasingly
higher \buoy\ values and larger offsets from the
control sample as one travels from north
(Line~56.0) to south (Line~90.0) down the
California coast.
}
\label{fig:N_cdfs}
\end{figure}

In addition to elevated \doxy,  
the representative hyperoxic \events\ along 
Line~90.0 also 
the hyperoxic parcels show a high buoyancy 
frequency (frequently $\buoy >15\,\nunits$).
This suggests that stratification may be a 
physical factor driving higher \doxy\
along Line~90.0.  
To explore this hypothesis further, 
Figure~\ref{fig:N_cdfs} compares the CDFs of \buoy\ for all
$\so > \exso$ parcels against the same control sample
as described above for \doxy.
The results for Line~90.0
confirm that high buoyancy frequency is a predominant trait
of hyperoxic water near the coast of San Diego.
Physically, 
stratification serves to
isolate these waters
from the surface to prevent air-sea fluxes 
from driving \doxy\ down to the 
maximum allowed oxygen concentration given its
temperature (i.e.\ $\so \approx 1$).

\begin{figure}[bt!] 
\centering
\includegraphics[width=0.95\linewidth, draft=false]{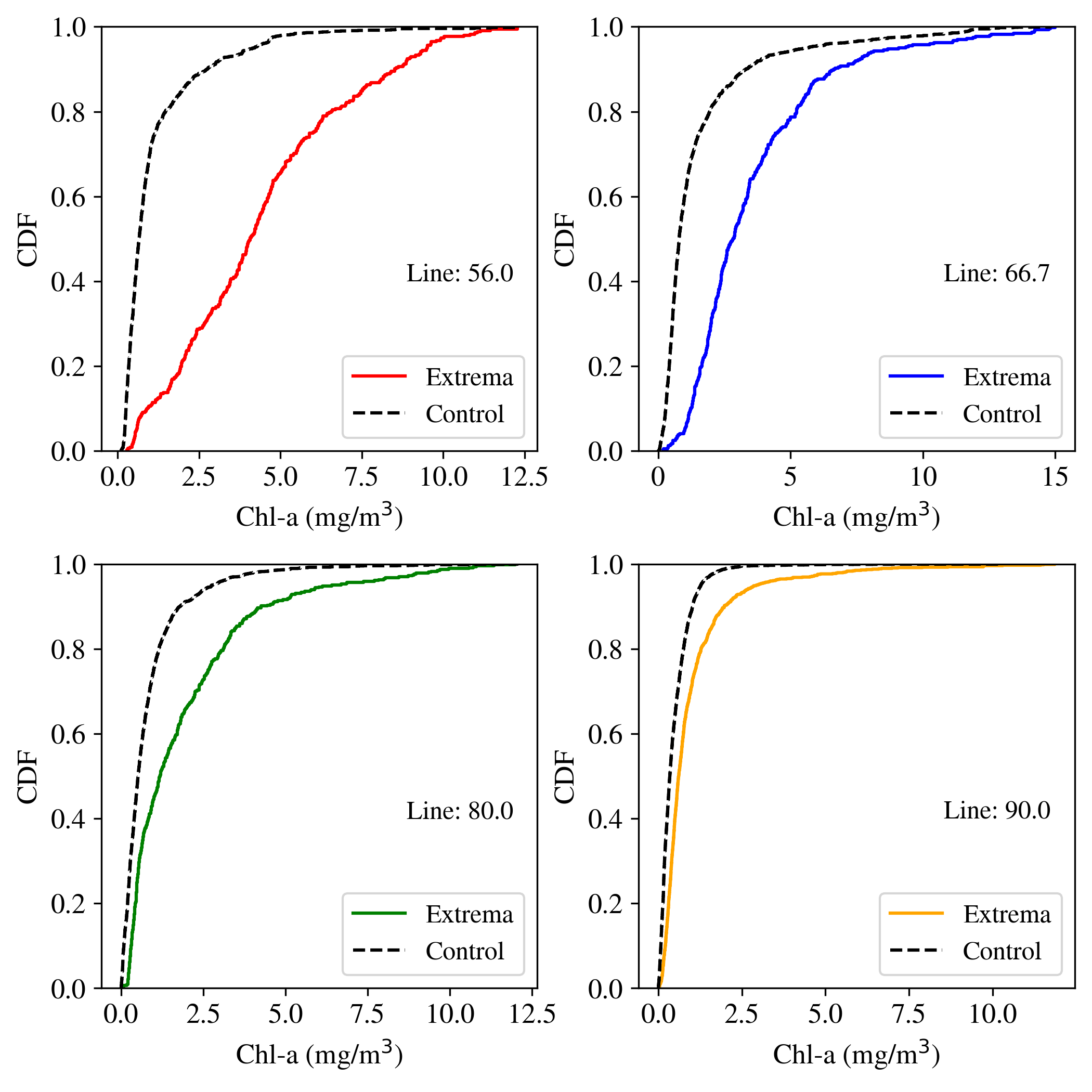}
\caption{Similar to Figure~\ref{fig:DO_cdfs}
but for \chla\ concentration.
Opposite to the results to for
buoyancy frequency, we find the extrema
nearly match the control sample along
Line~90.0 and show higher concentrations
and greater offset as one travels from
south to north along the coast.
}
\label{fig:Chl_cdfs}
\end{figure}

We also recall from Figures~\ref{fig:ex_90A} and
\ref{fig:ex_90B}, however, that a majority of the water
exhibits $\buoy > 10 \, \nunits$ during those events,
including parcels with $\so < \exso$ (although still
saturated).
The implication is that stratification is a necessary, albeit
insufficient property to achieve the highest saturation levels.
We further recall in these \events\ that the \chla\ concentration,
which often indicates enhanced productivity, does
not exhibit the highest values of the \event.
The values do generally exceed $0.2 \, \chlunits$ but are not extrema.
This point is statistically emphasized in 
Figure~\ref{fig:Chl_cdfs} which shows the CDFs
of \chla\ for hyperoxic extrema on Line~90.0 are 
nearly consistent with the control sample.
We hypothesize an additional factor not captured
by sensors on the \cugn\ gliders
-- perhaps enhanced nutrient content 
from recent upwelling -- 
is at work.


Consider next Line~56.0, the northernmost part of
the \cugn.  Referring to Figures~\ref{fig:N_cdfs},
we find the buoyancy frequency
CDF for the \so\ extrema
is consistent with the control sample.
We therefore conclude that stratification is not a primary factor
and further note that nearly all of the extrema occur at the
shallowest depth, i.e.\ in or just below the mixed layer.
On the other hand, the \chla\ CDF for Line~56.0
is greatly skewed to higher concentrations (Figure~\ref{fig:Chl_cdfs}).
For hyperoxic water to occur on this line, 
an intense bloom of productivity may be required.
In contrast with Line~90.0, we hypothesize the weaker stratification
implies a higher flux at the air-sea interface
which drives the oxygen saturation to 100\%\ unless
the \doxy\ generation greatly exceeds this flux.

Turning to the other Lines, 
these tend towards lower buoyancy frequencies than Line~90.0
(but still elevated).
This is well described by the CDFs in Figure~\ref{fig:N_cdfs}. 
Furthermore, as one travels North along the California coast, 
the \chla\ concentration of the extrema increases, 
both in absolute terms and relative to the control sample (Figure~\ref{fig:Chl_cdfs}).
In essence, the characteristics of the hyperoxic extrema on 
Lines 66.7 and 80 lie intermediate to the extrema on Lines 56.0 and 90.0.
We expect, therefore, that their hyperoxic \events\ 
are due to milder blooms in more stratified (i.e.\ less ventilated) 
waters.

\subsection{Origin of Hypoxic Extrema}

We comment briefly on the origin(s) of the near-surface,
hypoxic extrema in the \cugn.  Figure~\ref{fig:geo_low}
shows that these are preferentially within
100\,km of shore for Lines 56.0, 67.5, and 80.0
and at both $< 100$\,km and $\approx 150$\,km
on Line~90.0.  
These geographic regions are the primary locations of 
upwelling along each line, including the secondary
peak at $\approx 150$\,km on Line~90.0 due to the 
Santa Rosa Ridge.
Furthermore, except for Line~56.0,
the majority of these extrema occur during spring
or early summer (DOY=50-200) which are the most
common times for the wind-driven
upwelling of the \cugn\ system.

We note further 
that these extrema have temperatures lower than
the annual mean at their location and depth
(Figure~\ref{fig:Tanom_90}).
This is yet another indication that upwelling
drives these \events.
In addition,
the trend towards a higher incidence of
under-saturated \so\ as one travels north
(Figure~\ref{fig:SO_CDF}) is consistent
with the hypothesis.
Last, the modest but non-zero
buoyancy frequency values 
($\buoy \approx 5 \, \nunits$;
Figure~\ref{fig:SO_N})
imply the water has not mixed fully with
the surface.
We conclude that the near-surface,
under-saturated waters in the \cugn\ 
arise primarily from low-\doxy\ waters
during upwelling episodes.

\subsection{Previous Studies in the California Current System}

Previous studies on extrema of dissolved oxygen
in the California current system have focused on
hypoxic events \citep[e.g.][]{grantham2004,chan2008}.
The most extreme of these have had significantly negative
impacts on the ecosystem.
The dominant driver of these episodes is believed to be the advection
of abnormally low dissolved oxygen into the system
in combination with the upwelling of nutrient-rich but
oxygen-depleted water.
Furthermore,  hypoxia in the upper waters during these \events\ may be
linked to sustained stratification that prevents or reduces ventilation
\cite{bograd2008}.
Ironically, we have found stratification can play the same role
(reduced ventilation) but with the opposite outcome -- 
hyperoxic \events\ along Line~90.0.
An accurate estimate of the physical state of the California Current system,
therefore, is central to predicting biological processes.

%

\section{Conclusions}


This study has characterized the extremes of dissolved oxygen saturation in the California Current System using data from the California Underwater Glider Network. Hyperoxic events, defined as water parcels with oxygen saturation exceeding 110\%, were found to occur primarily within 100 km of shore, in the upper 30\,m of the water column, and during non-winter months. The drivers of these events appear to vary along the coast. Along Line 90.0 near San Diego, hyperoxic extrema are associated with high stratification, which may isolate sub-surface waters and allow biological production to elevate dissolved oxygen levels above saturation. In contrast, along the northernmost Line 56.0, hyperoxic events occur primarily at the surface and are strongly correlated with elevated chlorophyll-a concentrations, suggesting they are driven by intense phytoplankton blooms. Lines 66.7 and 80.0 show intermediate characteristics, with hyperoxic events likely resulting from moderate blooms in somewhat stratified waters.

Hypoxic extrema, defined as near-surface waters with oxygen saturation below 90\%, were found to occur predominantly within 100\,km of shore and during spring or early summer. These events are consistent with coastal upwelling bringing low-oxygen waters to the surface. The incidence of these hypoxic events increases from south to north along the California coast, reflecting the stronger influence of upwelling in the northern part of the study region. These findings highlight the complex interplay between physical processes like stratification and upwelling, and biological processes such as primary production, in shaping the dissolved oxygen dynamics of the California Current System. Understanding these extremes and their drivers is crucial for predicting ecosystem responses to changing ocean conditions and for improving biogeochemical models of this important coastal system.

\clearpage
\acknowledgments
J.X.P. and D.L.R. acknowledge support by the Simons Foundation.

%
%
\datastatement
All of the data analyzed within are available as products of the CUGN
\cite{Rudnick2016}.  
All of the code used to perform analysis is available on GitHub
\citep{cugn_doi}.








%




\bibliographystyle{ametsocV6}
\bibliography{so_paper}

\end{document}